\documentclass{aa}
\usepackage{amsmath,graphicx,amssymb}
\usepackage[varg]{txfonts}
\usepackage{natbib}
\usepackage{xcolor}
\usepackage{media9}
\usepackage{mathrsfs}  
\usepackage[T1]{fontenc}
\definecolor{DarkBlue}{rgb}{0.0,0.1,0.4}
\definecolor{Red}{rgb}{0.6,0.2,0.1}
\definecolor{DarkRed}{rgb}{0.5,0.0,0.5}
\definecolor{Blue}{rgb}{0.0,0.0,1.0}
\definecolor{Zelinkava}{rgb}{0.2,0.5,0.2}
\definecolor{Pink}{rgb}{0.0,0.7,0.7}
\definecolor{White}{rgb}{1.0,1.0,1.0}
\usepackage{hyperref}
\bibpunct{(}{)}{;}{a}{}{,}

\newcommand{\msun}{\ensuremath{M_\odot}}
\newcommand{\rsun}{\ensuremath{R_\odot}}

\hypersetup{
    colorlinks=true,
    linkcolor=Blue,
    citecolor=Blue,
    filecolor=Blue,     
    urlcolor=Blue,
}

\renewcommand{\d}{\text{d}}

\begin{document}

\title{Supernova explosions interacting with aspherical circumstellar material: 
Implications for light curves, spectral line profiles, and polarization}

\author{P.~Kurf\"urst\inst{1,2} \and O.~Pejcha\inst{1} \and J.~Krti\v{c}ka\inst{2}}

\institute{Institute of Theoretical Physics, Faculty of Mathematics and Physics, 
Charles University, V Hole\v sovi\v ck\' ach 2, 180 00 Praha 8, Czech Republic\and
           Department of Theoretical Physics and Astrophysics,
           Masaryk University, Kotl\' a\v rsk\' a 2, 611 37 Brno, Czech Republic
          }

\date{Received}

\abstract{
Some supernova (SN) explosions show evidence for an interaction with a pre-existing 
nonspherically symmetric circumstellar medium (CSM) in their light curves, spectral 
line profiles, and polarization signatures. The origin of this aspherical CSM 
is unknown, but binary interactions have often been implicated. To better 
understand the connection with binary stars and to aid in the interpretation of observations, 
we performed two-dimensional axisymmetric hydrodynamic simulations where an expanding 
spherical SN ejecta initialized with realistic density and velocity profiles 
collide with various aspherical CSM distributions. We consider CSM in the form of 
a circumstellar disk, colliding wind shells in binary stars with different 
orientations and distances from the SN progenitor, and bipolar lobes 
representing a scaled down version of the Homunculus nebula of $\eta$~Car. 
We study how our simulations map onto observables, 
including approximate light curves, indicative spectral line profiles at late times, 
and estimates of a polarization signature. We find that the SN--CSM collision layer 
is composed of normal and oblique shocks, reflected waves, and other hydrodynamical 
phenomena that lead to acceleration and shear instabilities. As a result, 
the total shock heating power fluctuates in time, although the emerging 
light curve might be smooth if the shock interaction region is deeply embedded 
in the SN envelope. SNe with circumstellar disks or bipolar lobes exhibit 
late-time spectral line profiles that are symmetric with respect to the rest velocity 
and relatively high polarization. In contrast, SNe with colliding wind shells 
naturally lead to line profiles with asymmetric and time-evolving blue and 
red wings and low polarization. Given the high frequency of binaries among 
massive stars, the interaction of SN ejecta with a pre-existing colliding wind 
shell must occur and the observed signatures could be used to characterize 
the binary companion.
}

\titlerunning{Supernova explosions interacting with aspherical circumstellar material}

\authorrunning{P.~Kurf\"urst et al.}
\maketitle

\section{Introduction}
\label{intro}

\begin{figure}
\begin{center}
\centering\resizebox{1.\hsize}{!}{\includegraphics{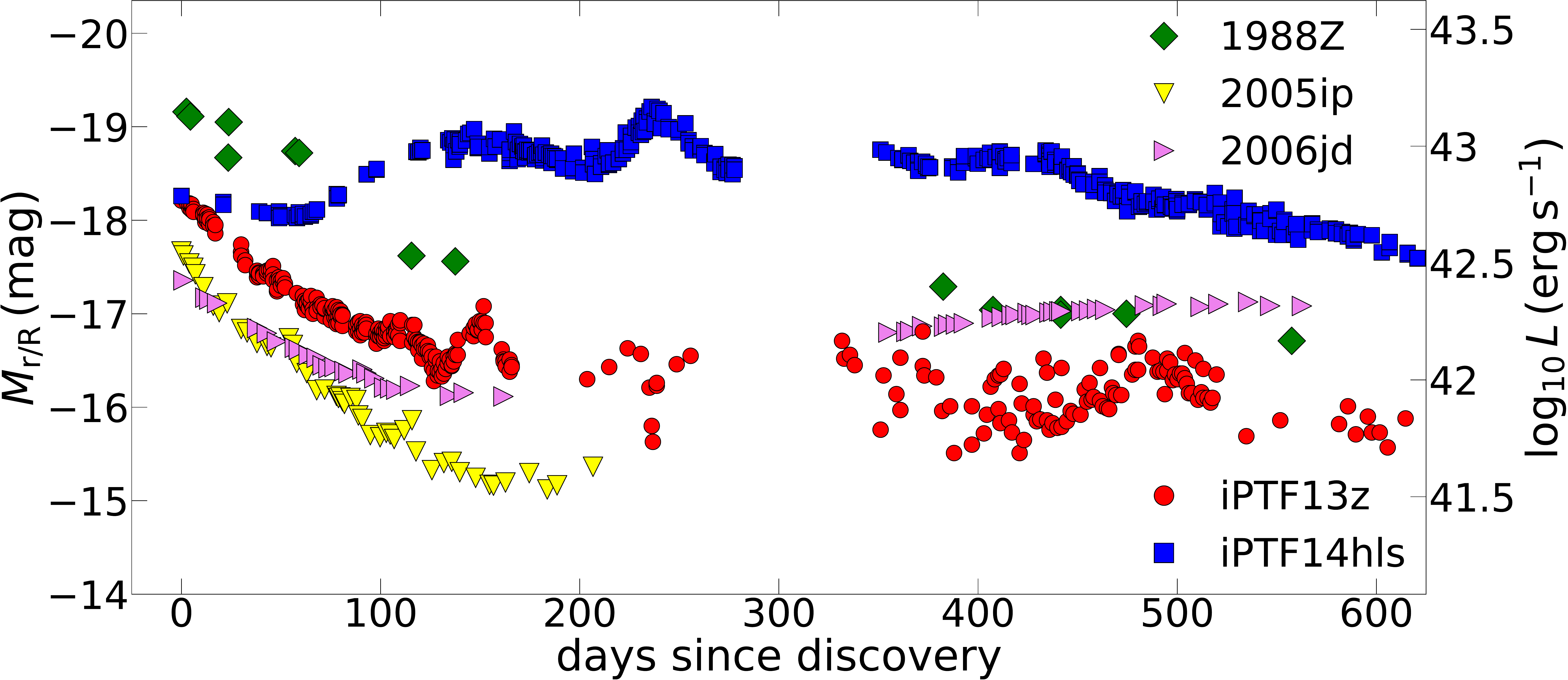}}
\caption{Comparison of light curves of five prominent long-lasting 
SNe II \citep[after][]{2017A&A...605A...6N}. The photometry data are 
from \citet{aretxaga99}, \citet{stritzinger12}, \citet{smith09}, 
\citet{2017A&A...605A...6N}, \citet{arcavi17}, and were obtained from the 
Open Supernova Catalog \citep{guillochon17}. We convert the $r/R$ magnitudes 
to bolometric $L$ according to Eq.~(1) in \citet{2017A&A...605A...6N}, 
which applies constant zero bolometric correction.}
\label{Nyholm} 
\end{center}
\end{figure}

When an expanding supernova (SN) blast wave collides with a dense pre-existing 
circumstellar material (CSM), the gas in the collision region is compressed and 
becomes radiative. Depending on the CSM properties, a substantial fraction of 
the SN kinetic energy might be converted into radiation. Such SN--CSM interactions 
can give rise to transients that are more luminous than ordinary SNe, including a subset 
of recently-recognized superluminous SNe \citep[e.g.,][]{2012Sci...337..927G,smith17_handbook}. 
We show light curves of a few examples of interacting SNe in Fig.~\ref{Nyholm}. 
Since the most radiatively efficient collisions occur with CSM located near the 
progenitor, the interacting SNe reveal the mass-loss history of massive stars 
shortly before the collapse of the core \citep[e.g.,][]{smith07,smith14,stritzinger12}.

The observed properties of the SN--CSM interaction often require an aspherical CSM 
distribution. The evidence comes from multicomponent line profiles in 
SN spectra \citep[e.g.,][]{chugai94,fransson02,smith15,andrews17,andrews18},  
(spectro)polarimetry \citep[e.g.,][]{leonard00,wang08,chornock10,patat11}, 
or combinations thereof \citep{bilinski18,bilinski20}. Aspherical CSM can lead 
to observable outcomes that are qualitatively different from a spherically symmetric SN--CSM 
interaction. If the shock interaction region subtends only a small fraction 
of the solid angle as seen from the SN, for example when the CSM is in the form 
of a disk, the SN ejecta can envelop and surround the shock and hide interaction 
signatures such as narrow emission lines. The SN can then behave as if there is 
an additional power source embedded deep in the ejecta. This qualitative picture 
can explain Type IIn-P SNe and similar objects \citep{mauerhan13,smith13}. 
A similar geometry of spherical explosion colliding with pre-existing 
equatorially-confined CSM could explain peculiar SNe, such as iPTF14hls \citep{arcavi17,andrews18} 
and AT2018cow \citep{margutti19}, transients associated with binary interactions and 
common envelope events \citep{metzger17,pejcha16a,pejcha16b,pejcha17,macleod18,hubova19}, 
classical novae \citep{li17}, and eruptions of very massive stars such as 
$\eta$~Car \citep{smith18}.

Aspherical CSM can have a number of different angular distributions. 
The commonly assumed configuration is an equatorially-confined disk with 
a radially decreasing density. This type of profile is a natural result of binary 
interactions such as mass transfer or common envelope 
\citep[e.g.,][]{podsiadlowski92,morris07,kashi10,smith11} as well as equatorial 
mass loss from rapidly spinning stars 
\citep[e.g.,][]{heger98,2001PASJ...53..119O,2007A&A...463..627K,2011A&A...527A..84K,
2014A&A...569A..23K,2018A&A...613A..75K}.

Other CSM distributions of interest are colliding winds within a binary star, 
where radiatively-cooled material from both winds accumulates in a dense curved 
surface resembling a bow shock \citep[e.g.,][]{stevens92,gayley97}. 
Since most massive stars are found in binaries \citep[e.g.,][]{sana12,moe17}, 
colliding wind shells must be relatively frequent around core-collapse SNe and 
might be confused with shells originating in pre-SN progenitor eruptions. 
Indeed, \citet{kochanek19} argue that a binary wind collision shell 
is a better explanation for the observed flash-ionized spectra of SN 2013fs. 
The bow shock-like structures resulting from wind collision shells are of 
particular interest to interacting SNe, because the collision with SN ejecta 
happens gradually and the interaction region moves progressively farther away 
from the progenitor star. Depending on the wind momenta of the binary components, 
the wind collision shell can be curved toward the progenitor, which would make 
the interaction signatures visible, or away from it, which could potentially 
hide the narrow lines from a large fraction of viewing angles. Furthermore, 
the wind collision shells in binary stars are corrugated due to hydrodynamical 
and radiation instabilities \citep[e.g.,][]{vishniac94,kee14,steinberg18}, 
which are a natural mechanism for forming clumps \citep{calderon16,calderon20}. 
The interaction of SN ejecta with clumpy CSM is expected to produce bumps in the 
light curves, which are seen in some events (Fig.~\ref{Nyholm}). Curved shells 
of dense material can also form when the binary companion does not have a strong wind, 
but it photoionizes a small HII region in the wind of the primary \citep{braun12,kochanek19}.  
Bow shock-like structures on larger scales also arise when red 
supergiant winds are trapped in a dense CSM shell by external ionizing 
photons \citep{mackey14}. 

Finally, the terminal SN explosion can collide with CSM that was shaped by 
preceding eruptions and their collisions. An example of such an event is the 
hypothetical future SN explosion of $\eta$~Car, which will sweep through the 
Homunculus nebula, which in itself might have been shaped by spherical 
eruption colliding with pre-existing equatorial outflow \citep{smith18}. 
Given the complex CSM pattern in this case, we expect a complicated behavior 
of multiple shocks and their mutual interactions.

Much of the theoretical work on observational characteristics of 
interacting SNe has been done in spherical symmetry 
\citep[e.g.,][]{chevalier82,moriya13,dessart15,dessart16}, but a few authors have 
explored SN interacting with aspherical CSM. \citet{blondin96} studied 
axisymmetric hydrodynamical simulations of a spherical self-similar driven 
wave propagating through a smooth angularly-dependent CSM and found that a 
protrusion develops in the directions where CSM is rarified. \citet{vanmarle10} 
performed two-dimensional hydrodynamic simulations of a collision of spherical 
SN ejecta with a spherical shell embedded in a spherical or an angularly-dependent wind. 
They estimated light curves by assuming optically thin cooling in the shock 
interaction region. \citet{vlasis16} utilized 2D hydrodynamic simulations with 
multigroup M1 radiation transport to calculate viewing angle-dependent 
light curves of various combinations of spherical and angularly-dependent 
SN ejecta and CSM, and a spherical SN colliding with a relic disk. \citet{mcdowell18} 
performed moving-mesh hydrodynamic calculations of spherical SN interacting with a disk. 
They estimated light curves by combining numerical shock heating rates with a 
diffusive model of SN light curves. \citet{kurfurst19} conducted two-dimensional 
hydrodynamic simulations of a spherical SN colliding with circumstellar disks of 
different masses embedded in spherically symmetric stellar winds with a range of 
mass-loss rates. Most recently, \citet{suzuki19} investigated interaction of 
a spherical SN ejecta with a circumstellar disk with axisymmetric special-relativistic 
adaptive mesh refinement hydrodynamics with two-temperature treatment of radiation. 
They calculated bolometric light curves from different viewing angles and estimated 
photosphere locations.

Here, we perform high-resolution axisymmetric hydrodynamic-only simulations of a 
spherical SN interacting with several aspherical CSM geometries. Our goal is to 
identify observable signatures that would allow us to discern different CSM geometries 
from basic SN observables.  Since many of the aspherical CSM distributions are 
linked to various aspects of binary star evolution, constraining the CSM geometry 
can probe the configuration of the binary. In particular, for the first time we 
study the hydrodynamical interaction of SN ejecta with colliding wind shells 
of binary stars. Since our goal is to illuminate qualitative differences 
between various CSM geometries, we do not attempt to provide models specifically 
matched to individual SN events.

This paper is organized as follows. Section~\ref{numset} describes the numerical 
setup of our calculations. Section~\ref{overhydro} describes and compares the 
hydrodynamic evolutions and discusses the shock interactions. 
In Section~\ref{sec:implications}, we provide approximate estimates of observable 
quantities based on our models (light curves, spectral line profiles, polarization) 
and compare these results to the observed SNe. In Sect.~\ref{conclude}, 
we summarize our findings.

\section{Numerical setup}
\label{numset}
In this section, we briefly review our numerical code (Sect.~\ref{sec:code}) 
and describe the initial and boundary conditions for our numerical experiments 
(Sect.~\ref{sec:initial_cond}).

\subsection{Description of the code}
\label{sec:code}
We conduct the numerical experiments using our Eulerian hydrodynamic code described 
in detail by \citet{2014A&A...569A..23K,2018A&A...613A..75K} and 
\citet{KKapplmath,kurfurst19}. We solve axisymmetric conservative equations of 
continuity, radial and polar components of momentum, and total energy on a polar 
grid described by radial coordinate $r$ and polar coordinate $\theta$. The equations are
\begin{align}
\label{conservcon01}
&\frac{\partial\rho}{\partial t}+ \frac{1}{r^2}\frac{\partial\widetilde M_r}{\partial r}+
\frac{1}{r\sin\theta}  \frac{\partial\widetilde M_\theta}{\partial\theta}=0,\\
[3pt] \label{conservativemomr}
&\frac{\partial M_r}{\partial t}+\frac{1}{r^2}\frac{\partial}{\partial r} 
\left(\widetilde M_r u_r\right)+ \frac{1}{r\sin\theta}\frac{\partial}{\partial\theta}
(\widetilde M_\theta u_r) -\rho\frac{u_\theta^2}{r}+ \frac{\partial P}{\partial r} = 0,\\
[3pt]\label{conservativemomtheta}
&\frac{\partial M_\theta}{\partial t}+\frac{1}{r^2}\frac{\partial}{\partial r} 
\left(\widetilde M_ru_\theta\right)+\! \frac{1}{r\sin\theta}\frac{\partial}{\partial\theta}
(\widetilde M_\theta u_\theta) \!+\!\rho\frac{u_r u_\theta}{r}\!+\!\frac{1}{r}\frac{\partial P}
{\partial\theta}=0,\\[3pt] \label{conservativeenerg}
&\frac{\partial E}{\partial t}+\frac{1}{r^2}\frac{\partial}{\partial r}
\left(\widetilde M_r H\right)+ \frac{1}{r\sin\theta}\frac{\partial}{\partial\theta}
\left(\widetilde M_\theta H\right)= 0,
\end{align}
where $\rho$ is the density, $u_r$ and $u_\theta$ are the radial and polar velocity 
components of the total velocity $u = \sqrt{u_r^2 + u_\theta^2}$, $P$ 
is the scalar pressure, $\widetilde M_r=r^2M_R$ and $\widetilde M_\theta=\sin\theta\ M_\theta$ 
are the two components of the momentum density, $E=\rho\epsilon+\rho u^2/2$ 
is the total energy density, $\rho\epsilon$ is the internal energy density, 
and $H=(E+P)/\rho$ is the enthalpy. The entire set of hydrodynamic equations 
is closed by an equation of state of a radiation dominated gas,  $3P=\rho\epsilon$. 
We estimate the adiabatic temperature in the models as $T=(3P/a)^{1/4}$, where $a$ 
is the radiation density constant. We estimate the specific entropy of the 
radiation dominated gas as $s=4aT^3/(3\rho)$. We perform the calculations in 
polar coordinates $(r,\theta)$, but it is often advantageous to present the 
results in cylindrical coordinates $\varpi=r\sin\theta$ and $z=r\cos\theta$.

We do not include gravitational forces, because we start our simulations when the 
fluid velocities are much faster than the local escape velocity, 
which makes the gravitational forces negligible. Since our focus is primarily 
on hydrodynamic interactions of the shocks, we do not include explicit viscosity, 
radioactive heating of the material, nor other effectively internal sources of energy 
like magnetar spin down. We do not include radiative cooling or any other 
redistribution of energy due to radiation, which corresponds to an assumption 
that the interaction region is optically thick and the diffusion time is longer 
than the expansion time. As the SN expands and the ejecta rarefies, 
this assumption will be violated. While simulations including radiation exist 
\citep{vlasis16,suzuki19}, proper treatment of radiation is beyond the scope of our work.

\subsection{Initial conditions}
\label{sec:initial_cond}

\begin{figure}
    \centering
    \includegraphics[width=0.435\textwidth]{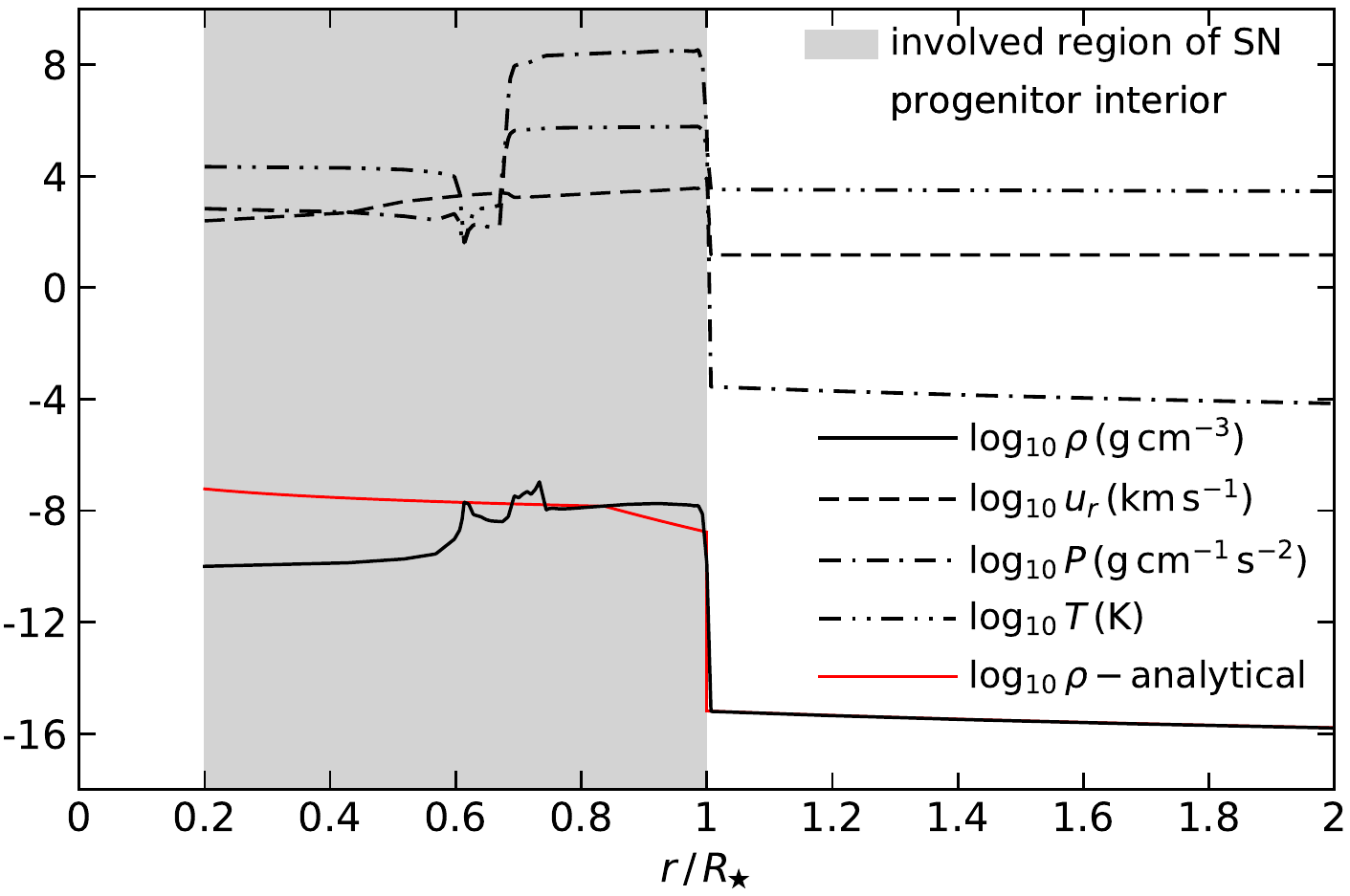}
    \caption{Initial profiles of selected hydrodynamic quantities. Black lines show the initial numerical profiles of density $\rho$, 
    radial velocity $u_r$, pressure $P$, and temperature $T$ of the SN ejecta 
    (gray region with $0.2 \le r/R_\star \le 1$) and of the stellar wind in the 
    close vicinity ($r > R_\star$). The red line shows the initial analytical 
    density profile described in Appendix~\ref{analsol}.}
    \label{fig:init_state}
\end{figure}
Our simulations include three components: spherically symmetric SN ejecta 
(Sect.~\ref{sec:sn_ejecta}), spherically symmetric stellar wind surrounding the 
progenitor star (Sect.~\ref{sec:wind}), and aspherical CSM. The properties of SN ejecta 
and spherical wind are the same between different simulations, but we study different 
types of aspherical CSM: equatorial disk (model A; Sect.~\ref{sec:disk}), 
colliding wind shell (models B1, B2a, B2b, and B3; Sect.~\ref{sec:coll_wind}), 
and bipolar lobes similar to the Homunculus nebula (model C; Sect.~\ref{sec:homunculus}). 
We describe the construction of the initial conditions of each of our 
models below in this section.

Our numerical grid covers $0.2 \le r/R_\star \le 450$, where $R_\star$ is the initial 
radius of SN ejecta, which corresponds to the precollapse radius of the progenitor star. 
The radial grid is composed of $60$ zones covering the initial SN ejecta for 
$0.2 \le r/R_\star \le 1$ and $6000$ zones between $R_\star$ and the outer boundary. 
The polar grid is uniform across the radial domain with $480$ grid cells for simulations 
covering $0 \le \theta \le \pi/2$ and $640$ cells for simulations covering 
$0 \le \theta \le 2\pi/3$. The  grid aspect ratio with much denser radial grid 
than the polar grid contributes to numerical stability by damping the so-called 
carbuncle instabilities that may develop near shocks  
\citep[e.g.,][see also \citealt{2017ASPC..508...17K,KKapplmath,kurfurst19}] 
{2001JCoPh.166..271P}. 

The boundary conditions at the inner and outer boundaries of the computational 
domain are free (outflowing). The reflection boundary condition is applied at the symmetry axis. 
The other lateral boundary is either reflecting or outflowing depending on the nature 
of the CSM distribution. The values in the grid boundary and ghost zones are extrapolated 
from the inner mesh computational domain values with a zeroth-order extrapolation. 
We evolve the system for a physical time of approximately $400$ -- $500\,\text{d}$. 

Since hydrodynamical instabilities are expected to play a significant role in our 
setup and we do not regularize our simulations by imposing artificial viscosity, 
we expect that local details of our results will depend on the numerical resolution. 
To test that the global properties are not sensitive to the resolution, we performed 
simulations of model A (Sect.~\ref{sec:disk}) with half the resolution in radial 
and polar directions outside of the progenitor. We plot light curves and indicative 
spectral line profile from this low-resolution run in Figs.~\ref{fig:lc} 
and \ref{spectra}, and find very good agreement with the default high-resolution models.

\subsubsection{Supernova ejecta}
\label{sec:sn_ejecta}

The initial profile of SN ejecta is often approximated with a broken power-law 
\citep[e.g.,][]{chevalier89,mcdowell18}, but here we use a more realistic initial 
condition. First, we calculate shock propagation through a realistic progenitor 
using one-dimensional implicit radiation hydrodynamics code SNEC 
\citep{2015ApJ...814...63M}. We use the default progenitor supplied with the code, 
which is a nonrotating star with the initial mass of $15\,\msun$ evolved to 
the moment of core collapse with the code MESA \citep{paxton11,paxton13}. 
Detailed information on the evolution and parameters used in the MESA 
calculation are given in Sect.~3.2 of \citet{2015ApJ...814...63M}. At the collapse, 
the progenitor is a red supergiant with the mass of $12.283\,\msun$ and radius 
$R_\star \approx 7.23\times 10^{13}\,\text{cm} \approx 1000\,\rsun$. We explode the 
progenitor with a $10^{51}$\,erg thermal bomb and keep all of the parameters to their 
default values except for the amount of radioactive nickel, which we set to zero. 
When the SN shock breaks out of the stellar surface, we remap the density, velocity, 
temperature, and pressure profiles from SNEC to our code. We show the initial profiles 
of these quantities in Fig.~\ref{fig:init_state}. Rotation of the progenitor 
star might lead to aspherical SN ejecta, however, to isolate the effect of 
aspherical CSM, we do focus only on spherical SN ejecta in this work. The effect 
of aspherical SN ejecta was explored by \citet{vlasis16}.

\subsubsection{Spherical wind}
\label{sec:wind}
SN ejecta is surrounded by a spherically symmetric wind, which serves primarily as a 
filling medium to ensure stable numerics. The density and temperature are too 
low to significantly affect the observed evolution. The density $\rho_\text{wind}$ 
is set to
\begin{equation}
\rho_\text{wind} =\frac{\dot{M}_\text{wind}}{4\pi r^2\varv_\text{wind}} = 
\rho_{0,\text{wind}} \left(\frac{R_\star}{r}\right)^2, 
\label{windpower0} 
\end{equation}
where $\dot{M}_\text{wind}=10^{-6}\,\msun\,\text{yr}^{-1}$ is the mass-loss rate and the 
$\varv_\text{wind}=15\,\text{km}\,\text{s}^{-1}$ is the asymptotic wind velocity 
typical for red supergiants \citep[e.g.,][]{goldman17}. This choice implies the wind base 
density $\rho_{0,\text{wind}}\approx 6.5\times 10^{-16}\,\text{g}\,\text{cm}^{-3}$. 
Total mass of the spherical wind over the full three-dimensional domain corresponding 
to our grid is $7\times 10^{-4}\,\msun$. We set the initial stellar wind temperature 
profile to be decreasing as $T_\text{wind}=T_\star \left(R_\star/r\right)^{0.5}$, 
where $T_\star\approx 3300$\,K is the progenitor stellar effective temperature. 
This corresponds to optically thin wind at radiative equilibrium with the progenitor 
radiation. At outer regions, the spherical wind has a fixed temperature  
$T_\text{wind}=15$\,K. We set the initial wind pressure and sound speed profile 
using the solar metallicity ideal gas law and the wind density and temperature. 
We do not take into account the acceleration of the wind close to the progenitor 
(Fig.~\ref{fig:init_state}). However, this typically affects only the first few days 
of the SN light curve \citep{moriya18} and we focus our work on later times 
of the SN evolution.

\subsubsection{Model A -- equatorial disk}
\label{sec:disk}
We set the density distribution as a sum of a spherically symmetric wind and an 
equatorially-concentrated disk. We set the disk density $\rho_\text{disk}$ 
following \citet{2018A&A...613A..75K} and \citet{kurfurst19} as 
\begin{align}\label{rhocsm}
 \rho_\text{disk}=\rho_{0,\text{disk}}\left(\frac{R_\star}{r}\right)^{w}\, 
 \exp\left[\frac{2\left(\sin\theta-1\right)}{(H/\mathcal{R})^2}\right],
\end{align}
where $\rho_{0,\text{disk}}
\approx 5\times 10^{-14}$\,g\,cm$^{-3}$ 
is the mass density at the base of the disk midplane (close to the surface of the 
SN progenitor star), $w=2$ is the power-law index \citep[same as in][]{mcdowell18}, 
$H$ is the disk vertical scaleheight, and $\mathcal{R}$ is the radial distance measured 
in the disk equatorial plane. We define $H=c_S/\Omega$, 
where $\Omega=\sqrt{GM_\star/\mathcal{R}^3}$ is the disk Keplerian angular 
velocity and $c_S$ is set using the same assumptions as for the spherical 
wind described in Sect.~\ref{sec:disk}.  The total mass of the disk and underlying wind is 
$5.2\times 10^{-3}\,\msun$, 
which is about a factor of ten higher than of the spherical wind alone. 
The disk parameters were chosen to provide a nontrivial shock interaction that 
would fit on our numerical grid.

\subsubsection{Models B -- colliding wind shells}
\label{sec:coll_wind}
A collision of stellar winds in a binary forms an interface of shocked gas. 
If the wind collision shocks are radiative, the slab collapses to a thin shell. 
Similar shells can form at the interface of a photoionized region inside the 
stellar wind of the progenitor. The ionizing photons come either from a hot binary 
companion or the ambient medium. Each of these situations as well as specific 
values of stellar parameters like wind momenta result in a somewhat different 
geometry and density profile of the shell. In particular, there are three different 
orientations of the shell with respect to the SN progenitor: shell oriented toward 
the progenitor (model B1), away from it (model B3), or the intermediate case 
of a planar shell positioned off the SN progenitor (models B2a and B2b). 

To focus on the effect that these three orientations have on SN--CSM collisions, 
we utilize somewhat idealized initial conditions inspired by a bow shock around 
a star with a spherical wind moving through a homogeneous external medium. 
The analytical structure of such bow shock was calculated by \citet{1996ApJ...459L..31W}. 
The basic parameter is the standoff distance of the shell from the SN progenitor $z_0$, 
which depends on the wind mass-loss rate from the progenitor, wind velocity, 
and density and velocity of the ambient medium. We choose $z_0$ sufficiently close 
to the progenitor so that the SN ejecta reache the bow shock within a year of the explosion. 
For models B1 and B2a, we choose  $z_0 = 215\,R_\star$, while for B2b and B3 we choose 
$z_0 = 50\,R_\star$. The values of $z_0$ were chosen arbitrarily to ensure 
that the initial shock collision occurs either far or close to the progenitor 
while keeping most of the shock interaction on our numerical grid. 
The surface density of the bow shock shell is 
\begin{equation}
\label{bow4}
\sigma_\text{bow}= \frac{\rho_\text{ISM}\left[2\alpha z_0^2\left(1-\cos\theta\right)+
\tilde\varpi^2\right]^2} {2\varpi\sqrt{z_0^4\left(\theta-\sin\theta\cos\theta\right)^2 +
\left(\tilde\varpi^2-z_0^2\sin^2\theta\right)^2}},
\end{equation}
where $\tilde\varpi(\theta)=z_0\sqrt{3\left(1-\theta\cot\theta\right)}$ 
is the cylindrical distance of the bow shock shell from the symmetry axis, 
and $\alpha \approx 6.7$ is the ratio of the ambient medium velocity and wind velocity. 
Our choice of $\alpha$ is relatively arbitrary and corresponds to a star with wind 
velocity of 15\,km\,s$^{-1}$ and ambient medium velocity of 100\,km\,s$^{-1}$. 
This gives the surface density at the standoff distance 
$\sigma_\text{bow}(z=z_0, \theta\rightarrow 0) = 3\rho_\text{ISM} z_0 (1+\alpha)^2/4 
\approx 10^{-6}\,\text{g}\,\text{cm}^{-2}$ \citep[cf.][]{1996ApJ...459L..31W}. 
Because we did not simulate the formation of the bow shock but only adapt the 
initial state described by Eq.~\eqref{bow4}, the volume density of the matter contained 
within the bow shock (and therefore also the bow shock thickness) is a free parameter. 
We selected this to achieve a difference in density between the bow shock standoff point 
and the adjacent stellar wind density to be about $3$ orders of magnitude. The initial 
temperature of the bow shock material is $T_\text{bow}\propto \rho^{1/3}$.

On the SN side of the bow shock, the density and temperature distribution is assumed 
to be of the spherical wind described in Sect.~\ref{sec:wind}. On the opposite side, 
we set the density to a constant value $\rho_\text{ISM} = 10^{-23}$\,g\,cm$^{-3}$ 
for all B models. This choice is unphysical, but it makes the interaction with the 
medium behind the bow shock negligible and allows us to focus the discussion on the 
geometry of the shell rather than the exact density profile behind it. The temperature 
behind the bow shock is set to $20$\,K. The total pre-explosion CSM masses for models 
B1, B2a, B2b, and B3 are $1.3\times 10^{-3}$, $2.1\times 10^{-3}$, $8.1\times 10^{-4}$, 
and $1.8\times 10^{-3}\,\msun$, respectively. 

In constructing the initial density profile, we neglected the Coriolis force that breaks 
the axial symmetry of the colliding wind shell. The thin shells are not oriented along 
the polar grid, which gives rise to density fluctuations along the shell. Although 
these variations are purely numerical, we expect that the real colliding wind shells 
are corrugated due to radiation and hydrodynamical instabilities and will effectively 
also have denser and thinner regions \citep[e.g.,][]{steinberg18,calderon20}.

\subsubsection{Model C -- bipolar nebula}
\label{sec:homunculus}

To set up a bipolar nebula similar to the Homunculus, we adopt the parameters for 
five major components of the nebula, that is, the preoutburst wind, great eruption, 
first post-outburst wind, minor eruption, and final post-outburst wind using equations (1-7) 
and Table~1 of \citet{2010MNRAS.402.1141G}. The size of the Homunculus nebula is 
too large to have SN--CSM interaction within the first year of SN evolution. 
To make the bipolar nebula fit within our computational grid, we reduced the duration 
of the two mass ejections. As a result, the total mass of the nebula is also lower, 
$1.7\times 10^{-2}\,\msun$. We set the initial temperature structure of the underlying 
stellar wind similar to the circumstellar disk, while we set the temperature of 
the two expanding eruption rings using polytropic approximation and ideal gas law. 
We set temperature in the surrounding unperturbed medium to $20$\,K.

For completeness, we note that $\eta$\,Car, the central star of the Homunculus nebula, 
is not a red supergiant. As a result, the density and velocity structure of the SN ejecta 
would be different, which would lead to somewhat different course of shock interaction and 
different early light curve. 

\section{Hydrodynamics of the interaction}
\label{overhydro}
\begin{figure*}
\begin{center}
\includegraphics[width=13.8cm]{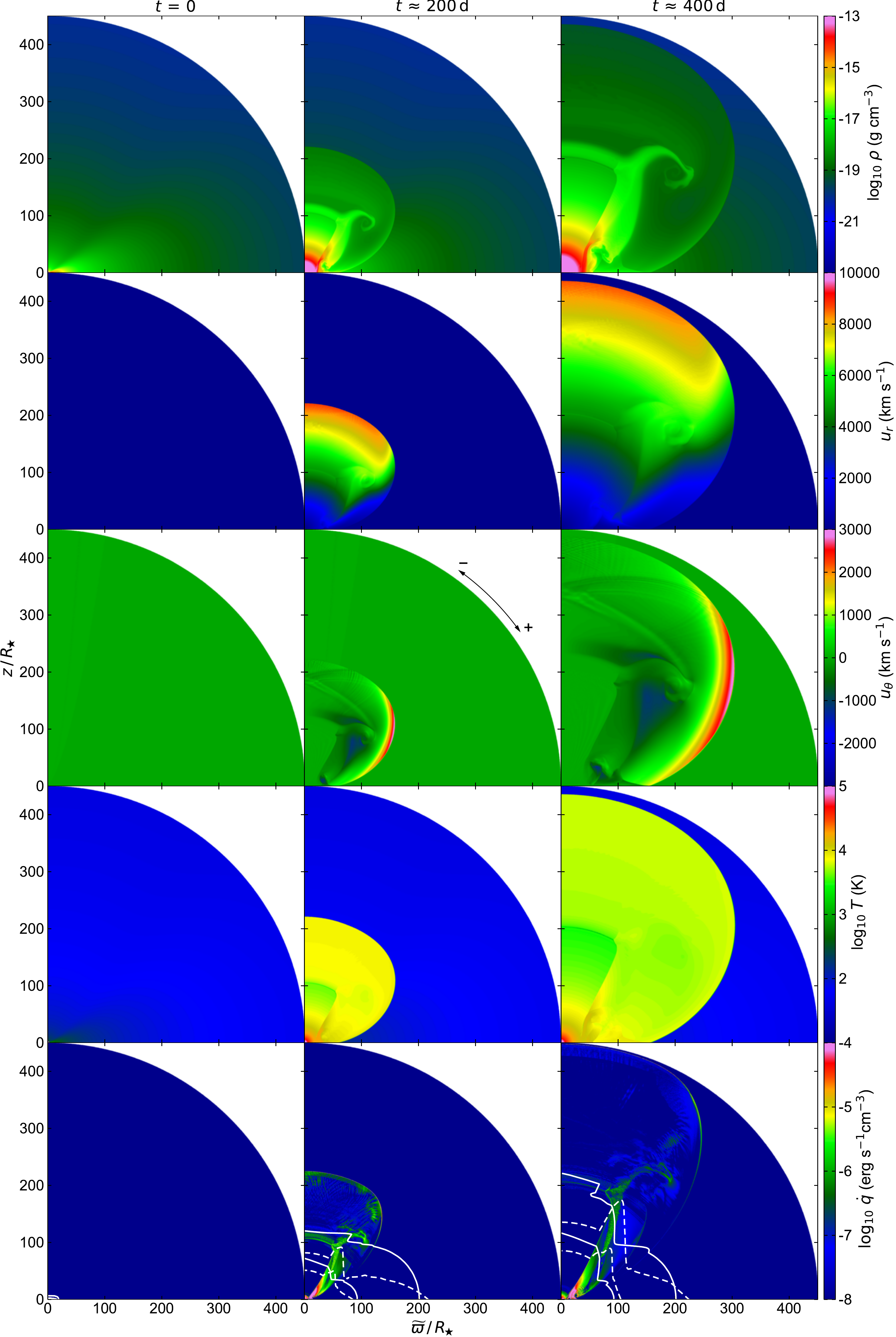}
\caption{Stages in the evolution of SN ejecta interacting with circumstellar disk 
(model A). The columns show snapshots at times $t=0$, $200$, and $400$\,days. 
Each row shows different quantity, from top to bottom: density $\rho$, radial velocity 
$u_r$, polar velocity $u_\theta$, temperature $T$, and shock heating rate $\dot{q}$. 
The bottom row also includes two contours of constant optical depth to electron scattering 
for two sight lines: $\theta = 0$ 
(looking from the top, dashed white lines)
and $\theta=\pi/2$ 
(looking from the right of the plot, solid white lines).
Inner and outer lines correspond to optical depths of $2/3$ and $0.1$, 
respectively. The $+$ and $-$ signs in the middle panel of the middle row illustrate 
the sign convention for polar velocity $u_\theta$. To conserve the size of the files, 
the resolution of the bitmaps was reduced down from the resolution of our simulations. 
Animated version of this figure is available as the~\href{run:movie_A.mp4}{movie A}.}
\label{fig2Ddisk} 
\end{center}
\end{figure*}

\begin{figure*}
\begin{center}
\includegraphics[width=13.8cm]{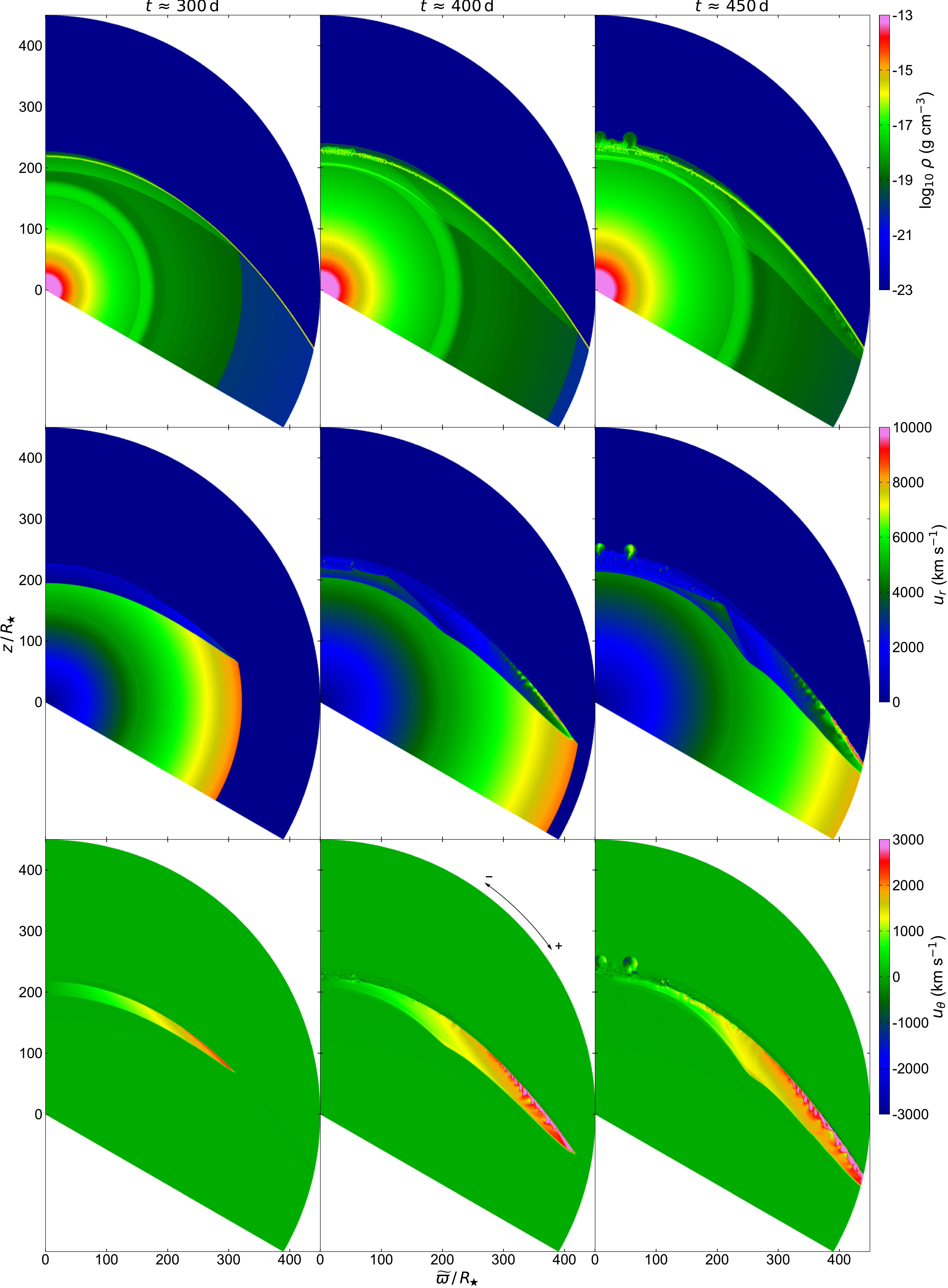} 
\caption{Stages in the evolution of SN ejecta interacting with a colliding wind shell 
oriented to the SN progenitor (model B1). The columns show snapshots at times 
$t=300$, $400$, and $450$\, days. Each row shows different quantity, from top to bottom: 
density $\rho$, radial velocity $u_r$, and polar velocity $u_\theta$. The remaining 
quantities are shown in Fig.~\ref{bowshockcolortiles_denser_rest}. 
Animated version of the two combined figures is available as 
the~\href{run:movie_B1.mp4}{movie B1}.}
\label{bowshockcolortiles_denser} 
\end{center}
\end{figure*}

\begin{figure*}
\begin{center}
\includegraphics[width=13.8cm]{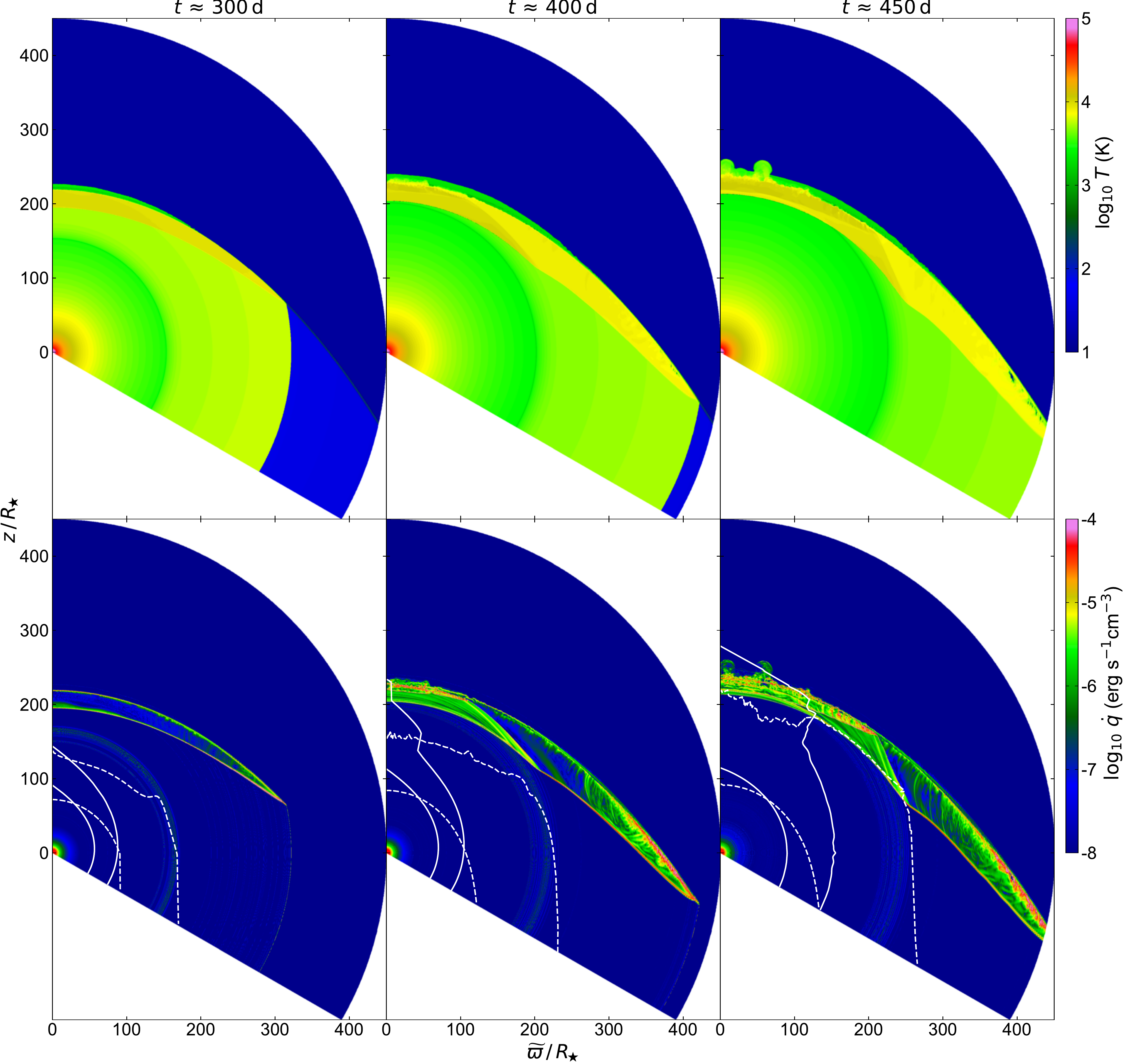}
\caption{Continuation of Fig.~\ref{bowshockcolortiles_denser}. The top row shows 
temperature $T$ and the bottom row displays shock heating rate $\dot{q}$. Similarly 
to Fig.~\ref{fig2Ddisk}, white lines in the bottom row correspond to contours of electron 
scattering optical depth, but evaluated from angles $\theta = 2\pi/3$ (solid lines) 
and $\theta=0$ (dashed lines).}
\label{bowshockcolortiles_denser_rest} 
\end{center}
\end{figure*}

\begin{figure*}
\begin{center}
\includegraphics[width=13.8cm]{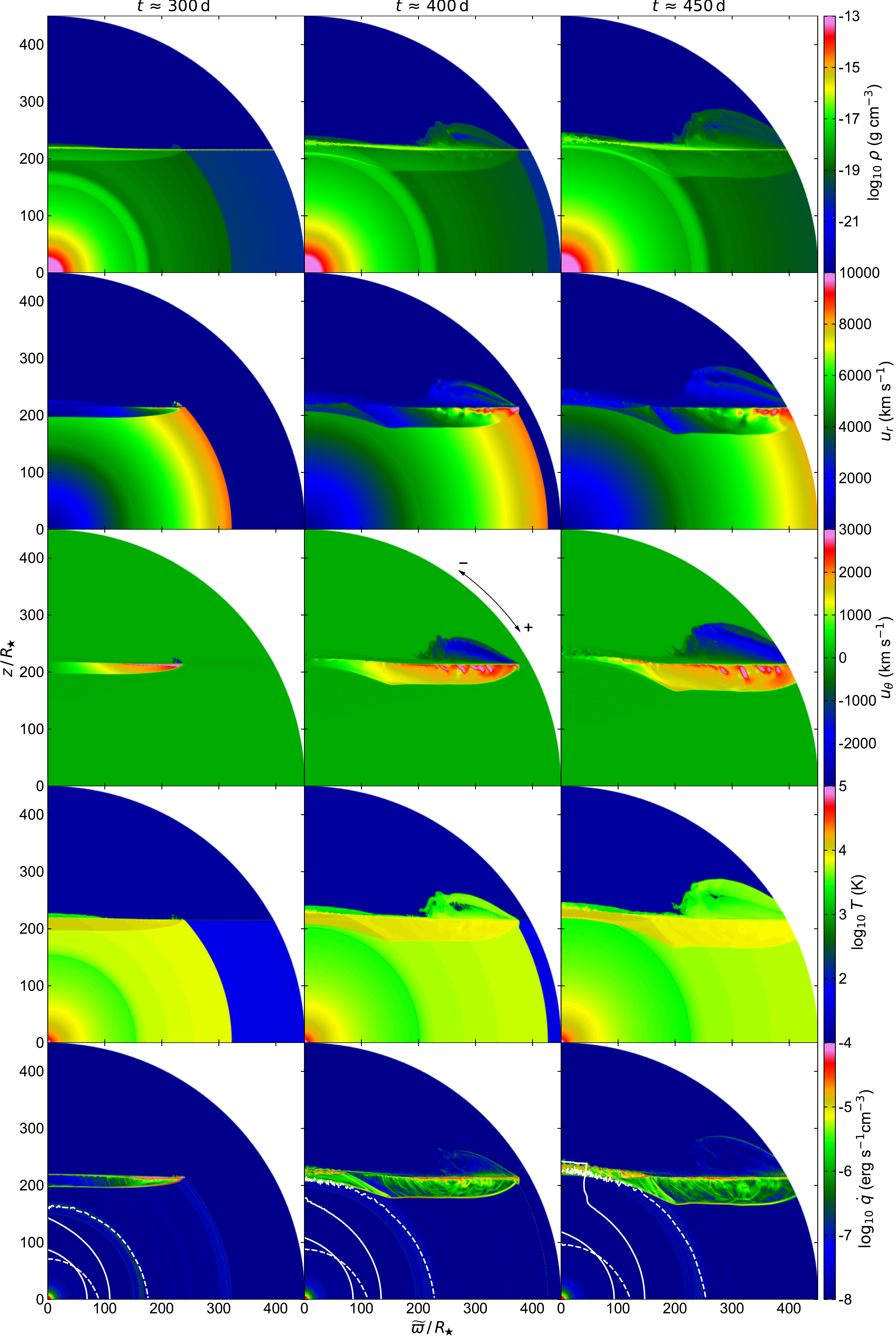}
\caption{Stages in the evolution of SN ejecta interacting with a planar shell located 
farther away from the progenitor (model B2a). The columns show snapshots at times 
$t=300$, $400$, and $450$\,days. The symbols and quantities are the same as in 
Fig.~\ref{fig2Ddisk}.
Animated version of this figure is available as the~\href{run:movie_B2a.mp4}{movie B2a}.}
\label{bowshock_denstraight} 
\end{center}
\end{figure*}

\begin{figure*}
\begin{center}
\includegraphics[width=13.8cm]{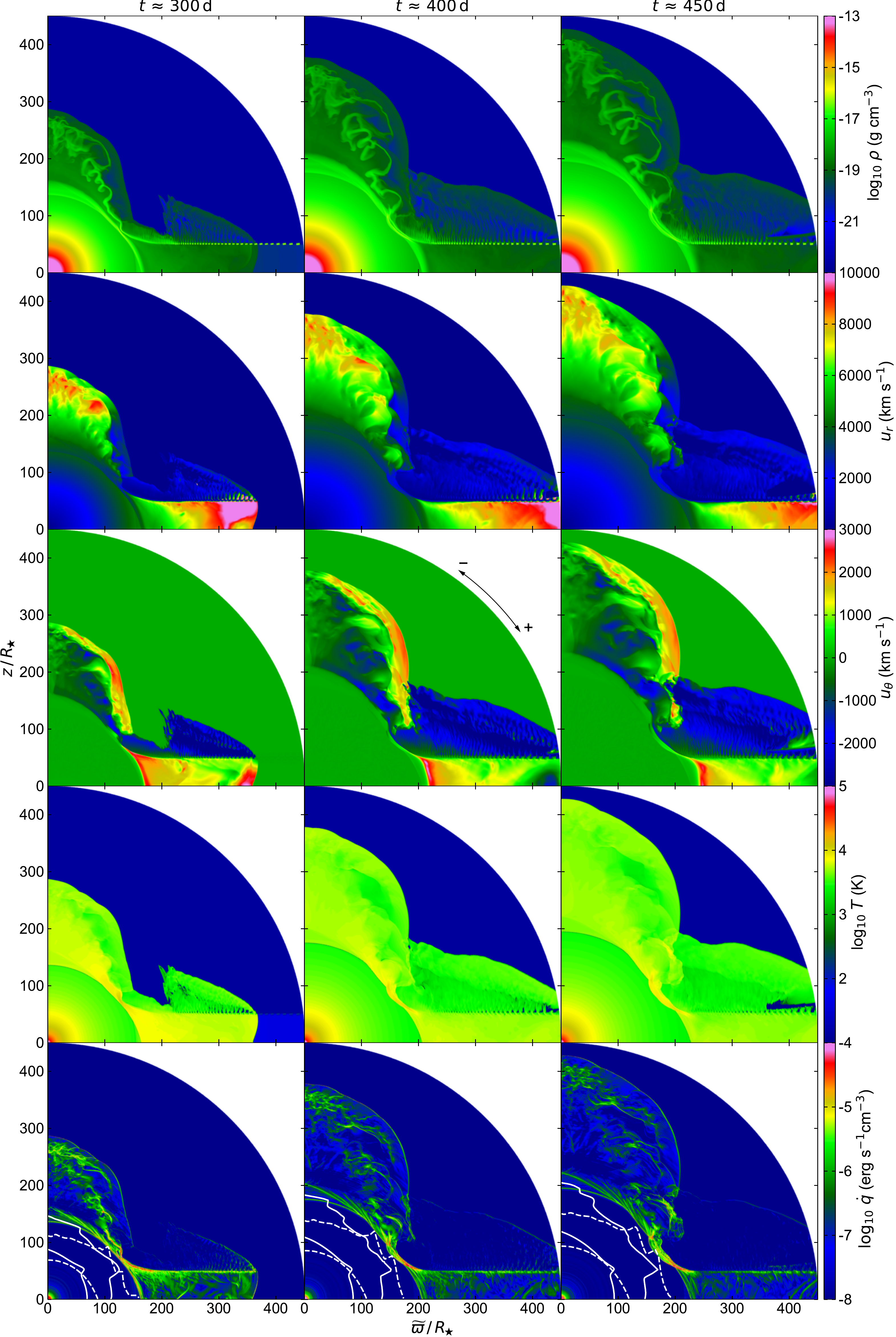}
\caption{Stages in the evolution of SN ejecta interacting with a planar shell located 
closer to the progenitor (model B2b). The columns show snapshots at times 
$t=300$, $400$, and $450$\,days. The symbols and quantities are the same as in 
Fig.~\ref{fig2Ddisk}. Animated version of this figure is available as 
the~\href{run:movie_B2b.mp4}{movie B2b}.}
\label{bowshock_denstraight_near} 
\end{center}
\end{figure*}

\begin{figure*}
\begin{center}
\includegraphics[width=13.8cm]{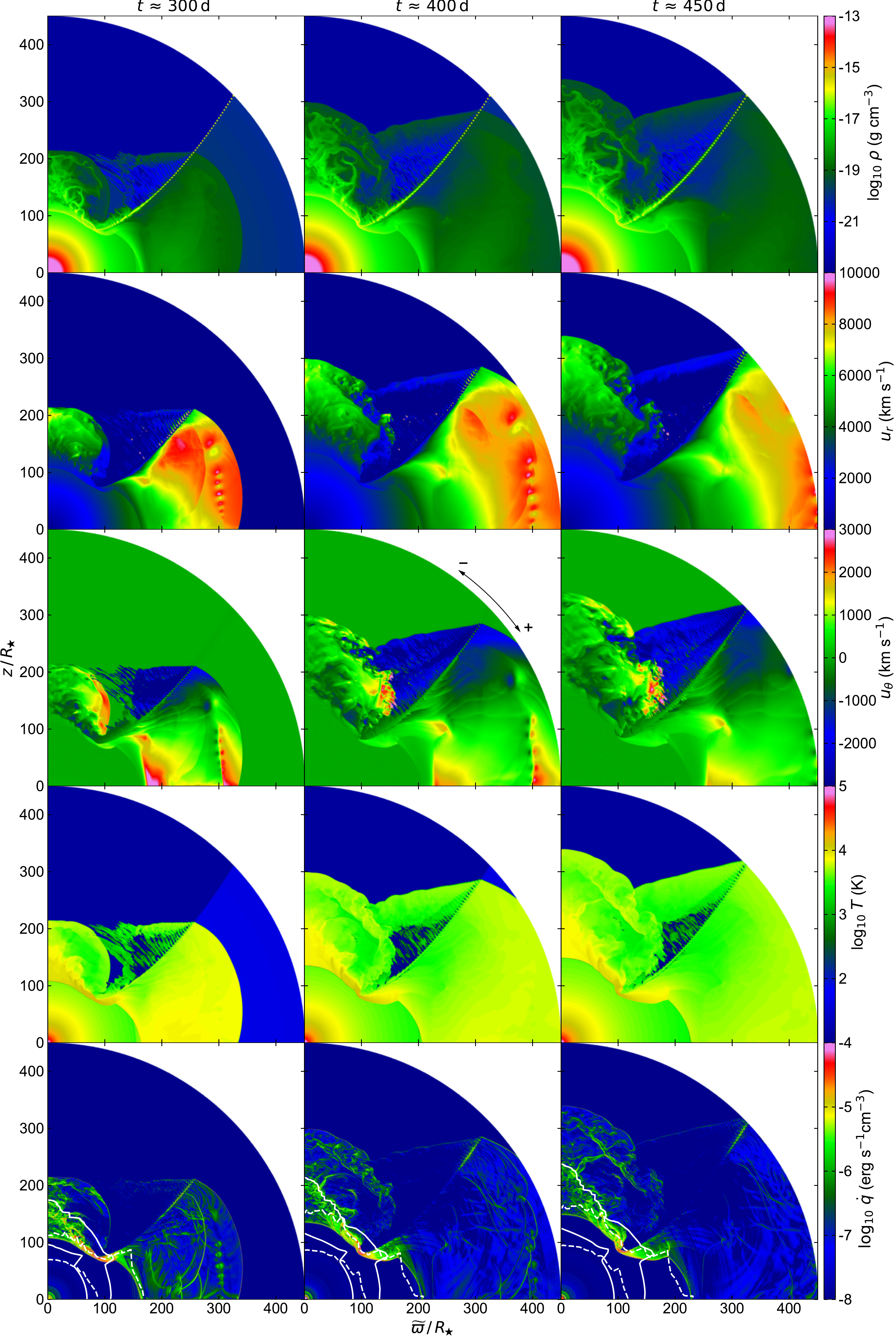}
\caption{Stages in the evolution of SN ejecta interacting with a colliding wind shell 
oriented away from the progenitor (model B3). The columns show snapshots at 
times $t=300$, $400$, and $450$\,days. The symbols and quantities are the same as 
in Fig.~\ref{fig2Ddisk}. Animated version of this figure is available as 
the~\href{run:movie_B3.mp4}{movie B3}.}
\label{bowshock_reversed} 
\end{center}
\end{figure*}

\begin{figure*}
\begin{center}
\includegraphics[width=13.8cm]{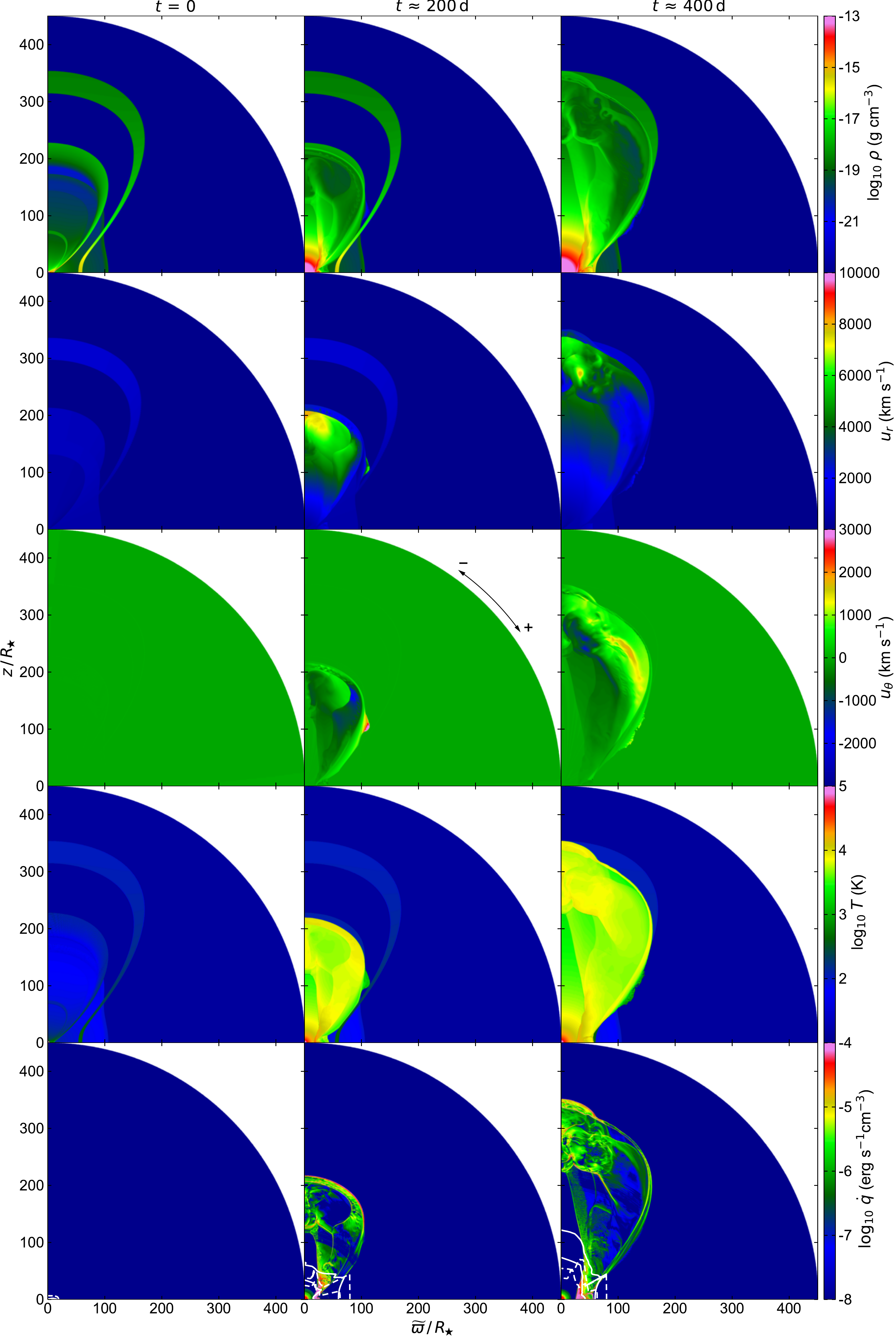} 
\caption{Stages in the evolution of SN ejecta interacting with bipolar lobes (model C). 
The columns show snapshots at times $t=0$, $200$, and $400$\,days. The symbols and 
the quantities are the same as in Fig.~\ref{fig2Ddisk}. Animated version of this figure 
is available as the~\href{run:movie_C.mp4}{movie C}.}
\label{lobescolortiles} 
\end{center}
\end{figure*}

In Figures~\ref{fig2Ddisk} -- \ref{lobescolortiles}, we present evolution of our models 
as tracked by density, radial and polar velocities, and temperature. For each model, 
we show the snapshots at three representative times corresponding to early, middle, 
and late stages of the interaction. To better visualize location and strength of shock 
interactions, we utilize the first law of thermodynamics and following \citet{mcdowell18}, 
we calculate the effective volumetric shock heating rate 
\begin{equation}
\dot{q} = \frac{1}{\gamma - 1}\frac{\text{d}}{\text{d}t}(P-P_\text{isen}),
\label{eq:qdot}
\end{equation}
where $\gamma = 4/3$ is the adiabatic index and $P_\text{isen}$ is the isentropic pressure 
corresponding to only adiabatic expansion or contraction. To calculate $P_\text{isen}$, 
we evolve the  constant $K=\rho^{-\gamma}P_\text{isen}$  within the hydrodynamic calculation 
as a passive scalar, evaluating $P_\text{isen}$ in each timestep. We show $\dot{q}$ 
in the bottom rows of Figs.~\ref{fig2Ddisk} -- \ref{lobescolortiles}.

\subsection{Circumstellar disk -- model A}
\label{numeromid}
The interaction of SN ejecta with circumstellar disk is shown in Fig.~\ref{fig2Ddisk}. 
This type of interaction was studied previously, although for slightly different initial 
density profiles of the disk \citep{vlasis16,mcdowell18,kurfurst19,suzuki19}. 
The results of our simulations are similar to previous works, but we review our results 
here as a baseline for an interpretation of more complicated models. When the SN ejecta 
collide with the disk, a localized shock is formed and travels outward through the disk. 
In our model A, the ratio of disk to ejecta mass is relatively small, and as a result, 
the SN ejecta do not become significantly decelerated and continue expanding also in the 
equatorial direction. Still, shearing motion between the freely expanding polar ejecta 
and the slower shock interaction region leads to development of Kelvin--Helmholtz instability, 
which we see as a prominent vortex in the density plots. It is possible that the 
vortex would be less prominent in three-dimensional calculations, where the turbulence 
cascade is inverse compared to two dimensions. The overpressure in the shocked gas pushes 
the material above and below the equatorial plane, which we see as a slight overdensity 
just above and below the shock interaction region with $u_\theta$ rapidly changing 
from negative to positive values. When the SN blast wave clears, these overdensities 
might be identified as expanding rings or cones. When comparing our results with previous 
works \citep{pejcha17,metzger17,mcdowell18,kurfurst19}, we note that the overdensity either 
leads or lags behind the equatorial shock, which suggests that the detailed behavior 
depends on initial conditions like the vertical and radial density profile of the disk.

\subsection{Colliding wind shells -- models B}
\label{bowshock}

We show the interaction of SN ejecta with a colliding wind shell oriented toward the progenitor 
(model B1) in Figs.~\ref{bowshockcolortiles_denser} and \ref{bowshockcolortiles_denser_rest}. 
This simulation was performed with a wider range of $\theta$ to see the shock interaction 
over a greater length of the shell. The interaction shock first appears at the standoff 
point on the vertical axis and then propagates away to greater $\theta$. The shell is 
first hit by the fast low-density SN ejecta, which do not have significant momentum 
to cause noticeable expansion along the $z$ axis. But eventually, the shell buckles 
and we see SN ejecta expanding above the shell in a series of Rayleigh--Taylor plumes. 
As the shock interaction spreads laterally to higher $\theta$, progressively larger 
fraction of the SN ejecta velocity is oriented along the shell rather than perpendicular 
to it. Consequently, SN ejecta interacting with the shell achieves large positive 
$u_\theta$, as seen in the bottom row of Fig.~\ref{bowshockcolortiles_denser}. 
We witness development of shearing instabilities along the shell, which is especially 
well seen at $t=450$\,days around $\varpi \approx 400 R_\star$ in the plots of 
$\rho$, $u_r$, and $u_\theta$. We expect that more realistic initial conditions 
with perturbations of the colliding wind shell \citep[e.g.,][]{calderon20} would 
lead to a faster development of the instabilities. Finally, the bottom row of 
Fig.~\ref{bowshockcolortiles_denser_rest} shows the volumetric shock heating 
rate $\dot{q}$. We can identify forward and reverse shocks that bound the banana-shaped 
shock interaction region. Between $t=300$ and $400$\,days the peak of $\dot{q}$ 
moves along the shell away from the axis, but at $t=450$\,days the standoff point 
is reached by denser parts of the SN ejecta and $\dot{q}$ significantly increases 
for $|\theta| \lesssim \pi/4$. 

In Figure~\ref{bowshock_denstraight}, we explore the hydrodynamics of SN ejecta colliding 
with a plane shell (model B2a) with a standoff point located at the same distance as 
in model B1. The development of the instabilities is similar to model B1 in the sense 
that Rayleigh--Taylor plumes appear above the shell close to the $z$ axis and 
Kelvin--Helmholtz vortices develop along the shell at larger $\varpi$. We also see a 
plume traveling above the shell back to the axis of symmetry with negative $u_\theta$, 
as shown in the middle row of Fig.~\ref{bowshock_denstraight}. This plume is caused 
by low-density regions in the initial density distribution of the shell, which is 
an artificial feature discussed in Sect.~\ref{sec:coll_wind}. This material has low 
density and does not show high $\dot{q}$, which implies that its observational 
consequences might be minor.

In Figure~\ref{bowshock_denstraight_near}, we show the same geometry configuration as 
in Figure~\ref{bowshock_denstraight}, but with a shell positioned significantly closer 
to the SN progenitor (model B2b). The evolution for the closer shell proceeds somewhat 
differently than for the shell positioned farther away. Due to the proximity of 
the shell to the SN progenitor, the high-momentum part of the ejecta hits the shell 
within the simulation time and the SN ejecta is able to break through the shell. We 
see that the denser parts of the SN ejecta become somewhat equatorially flattened, 
but the expansion generally continues in all directions. The interaction of the ejecta 
with the shell causes the formation of a thin filamentary overdensity, which winds in a 
complicated pattern in the outer regions of the polar ejecta. Similar but not identical 
filamentary pattern is seen also in the volumetric shock heating rate in the bottom row. 
We expect that in three-dimensional simulations or with corrugated initial density 
distribution of the shell, the filamentary structure would be less organized and perhaps 
completely dispersed. We expect that this region would still exhibit mixing and shock 
heating. Although there is noticeable shock heating around the filaments, the peak of 
$\dot{q}$ concentrates in areas where the shell is hit by SN ejecta with sufficiently 
high momentum to destroy the shell at that position. Finally, we expect that hydrodynamics 
in the case of a more distant shell (Fig.~\ref{bowshock_denstraight}) would eventually 
resemble the behavior seen in Fig.~\ref{bowshock_denstraight_near} if the simulation 
was followed to later times and over larger domain sizes.

In Figures~\ref{bowshock_reversed}, we illustrate the hydrodynamical evolution for 
a colliding wind shell oriented away from the SN progenitor (model B3). SN ejecta breaks 
through the shell near the standoff point on the $z$ axis and we witness instabilities 
and shock heating as the ejecta propagates in the polar direction, similarly to models 
B1 and B2b. At greater distances from the SN progenitor, the shell becomes almost parallel 
with the radial direction and the shock interaction does not occur there as the fast ejecta 
sweep around. As a result, $\dot{q}$ remains high only relatively close to the progenitor. 
Ultimately, the shell will be destroyed as the bulk of the shocked region accelerates and 
moves outward, but we do not see this happening within the duration of simulation of model B3.

Finally, we point out several differences between SN ejecta interaction with a circumstellar 
disk and a colliding wind shell. In the case of disk interaction, the shock was localized 
primarily in the disk and only moved radially due to its bulk motion. In other words, the 
system could be described with two components: freely expanding SN ejecta in the polar 
direction with relatively small perturbations such as the overdense shoulder above 
and below the shocked region and the associated Kelvin--Helmholtz vortex, and strong shock 
propagating in the equatorial plane. Each of these components could be easily approximated 
with a separate spherically symmetric calculation. Indeed, we verified that the evolution 
near the equatorial plane could be relatively well described by a SNEC simulation with 
spherically symmetric CSM density distribution. Interaction with the colliding wind shell 
is fundamentally different, because the shocked region moves not only due to its bulk motion, 
but also because different parts of the shell get hit by different parts of the SN ejecta 
at different times. Although this type of interaction could be described as a superposition 
of a large number of collisions at progressively increasing radii, our simulations reveal 
that hydrodynamical instabilities are more vigorous than for the disk and effectively 
couple together evolution at different locations on the shell. When viewed from the SN 
progenitor, colliding wind shells can subtend much larger fraction of the solid angle 
than a circumstellar disk, which means that proportionally larger fraction of SN ejecta 
eventually significantly interacts with the CSM.

\subsection{Bipolar nebula -- model C}
\label{bilobes}
In Figures~\ref{lobescolortiles}, we show the hydrodynamical evolution of SN ejecta 
colliding with a bipolar nebula modeled after the Homunculus, but with smaller spatial 
scale and lower total mass (model C).  We see that SN ejecta is deformed to resemble 
the original CSM shape. Since the lobes completely enclose the progenitor, the SN ejecta 
cannot pass around the denser parts of the CSM to expand freely in some direction 
(unlike the case of the circumstellar disk). As a result, we see reflected waves of 
the material propagating back into the regions within the original lobes, which is 
particularly noticeable in the plots of the density. Higher CSM density near the equator 
gives rise to a shocked region that is somewhat similar to the features seen in model A. 
Due to the complicated CSM geometry, various hydrodynamical instabilities are not as easy 
to localize as in the simpler geometries. 

\section{Implications for observations}
\label{sec:implications}
Different CSM configurations result in qualitatively different hydrodynamical behavior 
in our simulations. We are now interested in finding observables, or their combinations, 
that would allow us to distinguish between different CSM geometries. We provide estimates 
of observable signatures for light curves (Sect.~\ref{lightcurves}), late-time spectral 
line profiles (Sect.~\ref{lineprofiles}), and polarization (Sect.~\ref{polarization}). 
We discuss implications for some observed SNe in Sect.~\ref{compare}.

\subsection{Light curves}
\label{lightcurves}
\begin{figure*}
\centering
\includegraphics[width=0.9\textwidth]{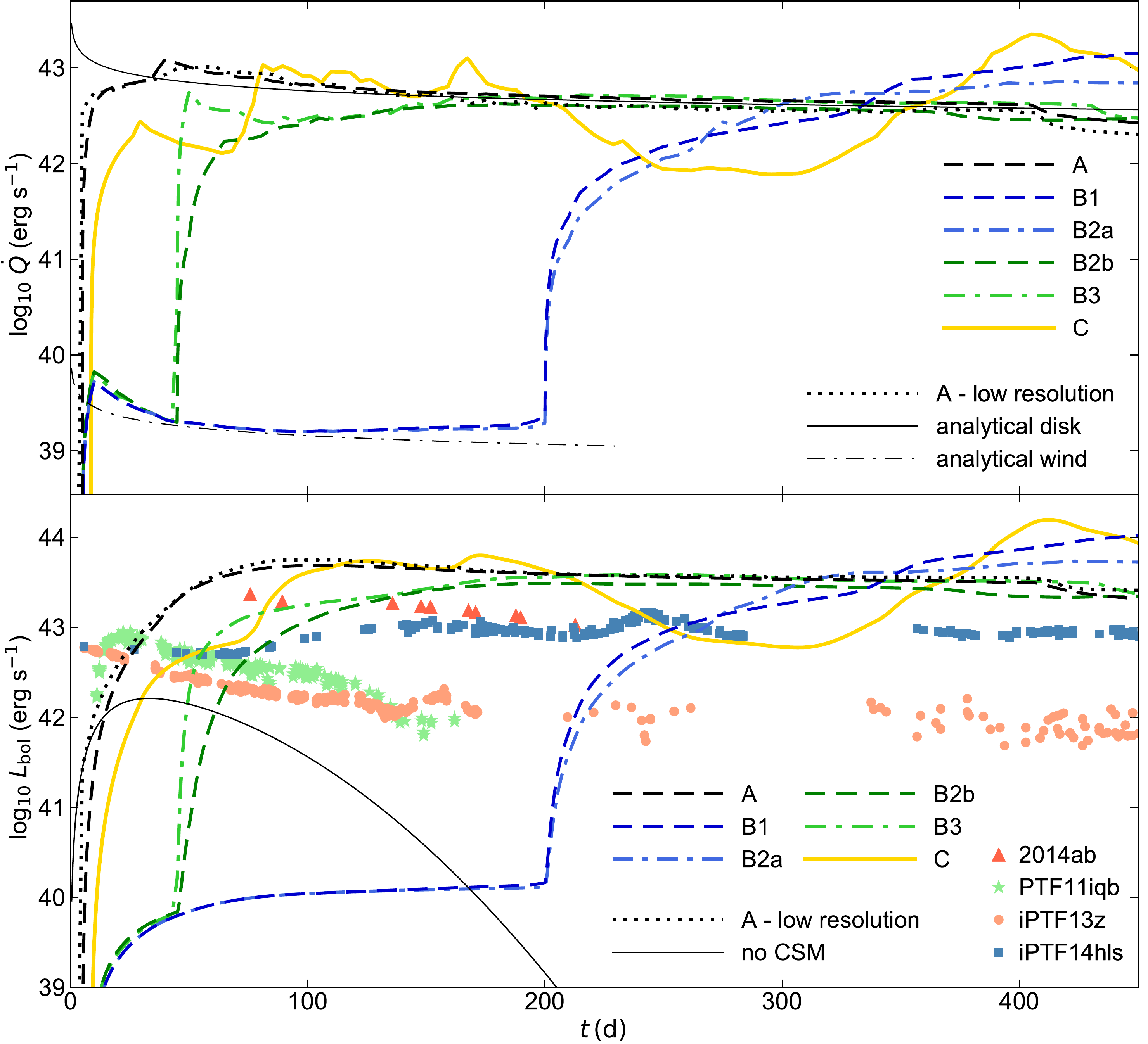}
\caption{Estimates of light curves from our simulations. \textit{Top}: 
Shock heating rate $\dot Q$ for our models (thick black dashed line for model 
A and thick colored lines for other models). The thin black dash-dotted 
line shows the analytic solution for shock passing through the wind with the same 
parameters as in Sect.~\ref{sec:wind}. The thin black solid line 
shows the analytic solution for interaction with a circumstellar disk  
(Eq.~[\ref{power5}], slope parameter $n=12$, disk opening angle $\widetilde H=0.2$).
The black dotted line demonstrates the $\dot Q$ for model A calculated with lower 
resolution.
\textit{Bottom}: Bolometric light curves from Eqs.~\eqref{lumin} and \eqref{lumin2}. 
The thick black and colored lines are estimated light curves for our models. 
The black dotted line shows the light curve for model A calculated with 
lower resolution. The thin black solid line labeled ``no CSM'' corresponds 
to a hypothetical SN without CSM interaction but heated with $0.28\,\msun$ of radioactive 
nickel. This choice roughly corresponds to the brightest normal hydrogen-rich 
SNe \citep{pejcha15,2017ApJ...841..127M} or the population mean of stripped-envelope 
SNe \citep{anderson19}. The calculated bolometric light curves are compared with 
the observed light curves of SN2014ab \citep{bilinski20}, PTF11iqb \citep{smith15}, 
iPTF13z \citep{2017A&A...605A...6N}, and iPTF14hls \citep{arcavi17} shown with points.}
\label{fig:lc} 
\end{figure*}

Without self-consistent treatment of radiation and hydrodynamics, we cannot make 
quantitative predictions for specific CSM properties, but we can investigate general 
trends using simpler semi-analytic models. In estimating the radiative luminosity, we must 
determine whether the optical depth to the radiative shock is relatively small and 
the instantaneous luminosity is proportional to the shock power, or whether the optical 
depth is large and the shock radiation diffuses out through a reprocessing layer. 
In bottom rows of Figs.~\ref{fig2Ddisk}--\ref{lobescolortiles}, we overplot on the 
volumetric shock heating rate contours of electron scattering optical depths from two 
different sightlines under the assumption that the region is completely ionized. 
If the majority of the shock power is contained within a contour, the diffusion 
approximation is more appropriate than optically-thin treatment. Our estimate of 
the photosphere is very rough because we ignore other temperature- and 
density-dependent sources of opacity. For example, radiative shock can keep its 
vicinity ionized for a longer time and thus prevent hydrogen recombination and a 
drop of optical depth 
\citep[e.g.,][]{smith13_etacar,smith17_handbook,smith15,metzger17,andrews18,margutti19}. 
We thus consider both possibilities that the shock power is either instantaneously 
converted to optical radiation (Sect.~\ref{shockpower}) or it diffuses out of optically-thick 
envelope (Sect.~\ref{simplelightcurves}).

\subsubsection{Shock power as a function of time}
\label{shockpower}
In the optically-thin case, the SN luminosity $L_\text{SN}$ is approximately 
proportional to the shock heating rate $\dot{Q}$ of the radiative shock, 
$L_\text{SN} \approx \dot{Q}$, and we thus discuss the behavior of $\dot{Q}$. 
We calculate $\dot{Q}$ numerically by integrating the volumetric shock heating rate 
$\dot{q}$ given by Equation~(\ref{eq:qdot}) over the simulation volume
\begin{equation}
\label{power6}
\dot{Q}=\int\dot{q}\,\text{d}V.
\end{equation}
Solid lines in the upper panel of Fig.~\ref{fig:lc} show the time evolution of 
$\dot{Q}$ in our simulations. The dashed line corresponds to analytical calculation 
of shock power in the case of shock interaction with stellar wind (lower line) 
and circumstellar disk (upper line). Details of the calculation are given in 
Appendix~\ref{analsol}. 

We see from Fig.~\ref{fig:lc} that $\dot{Q}$ tracks the CSM density encountered by 
the shock. Model A (circumstellar disk) exhibits gradual decrease in $\dot{Q}$, 
because the shock slows down as it sweeps up more mass of the disk. For models B1 
(colliding wind shell oriented to the SN progenitor) and B2a (distant planar shell) 
in the first $200$\,days, the shock heating rate stays  at 
$\approx 3\times 10^{39}$\,ergs\,s$^{-1}$, which is caused by the SN ejecta 
colliding with the spherically symmetric stellar wind. Later, $\dot{Q}$ increases 
gradually as a larger fraction of the SN ejecta interacts with the CSM. The rise in $\dot{Q}$ 
exhibits wiggles of $\lesssim 10\%$. For models B2b (close planar shell) and B3 
(colliding wind shell oriented away from the SN progenitor), we see a fast rise 
in $\dot{Q}$ as the dense parts of the SN ejecta hit the nearby CSM at early times, 
which is followed by a gradual decline as progressively more distant regions of CSM 
are shocked. These models also exhibit fluctuations on a similar level to B1 and B2a. 
We expect that the shock heating rates of models B could exhibit potentially larger 
fluctuations with more realistic initial conditions that take into account corrugation 
of the colliding wind shells. Model C (bipolar nebula) presents the most complicated 
behavior with several high-amplitude bumps and wiggles caused by multiple shells in the CSM. 

The shock power is also a useful quantity to compare numerical results with analytic 
estimates and with previous results on the same problem. Focusing on model A in the top panel 
of Fig.~\ref{fig:lc}, we see that for most of the time the numerical results 
closely match the analytical predictions elaborated in Appendix~\ref{analsol}. 
Apart from the initial transient, a small disagreement is seen at approximately 
$70$\,days when numerical results give higher $\dot{Q}$ than analytical estimates 
and after $400$\,days when the numerical $\dot{Q}$ is slightly lower than the 
analytical ones. Similar pattern is seen in Fig.~4 of \citet{mcdowell18}. 
This suggests that our results are in agreement with those of \citet{mcdowell18} 
and that the realistic initial profile of the SN ejecta has relatively small 
effect on the outcome compared to the broken power-law used in analytical estimates 
and by \citet{mcdowell18}.

\subsubsection{Shock power as an internal power source}
\label{simplelightcurves}
To estimate light curves under the assumption of shocks deeply embedded in the ejecta, 
we follow the semianalytic calculation of SN bolometric light curves 
introduced by \citet{1980ApJ...237..541A,1982ApJ...253..785A}. We adopt the diffusion 
timescale $\tau_0 \propto \kappa M_\text{SN}/R_\text{SN}$, where $\kappa$ is the 
(Thomson) opacity ($\kappa\approx 0.34\,\text{cm}^2\,\text{g}^{-1}$), $M_\text{SN}$ is 
the total mass of material that is involved in the explosion (ejecta + CSM), 
and $E_\text{SN}$ is the SN explosion energy, the hydrodynamical time 
$\tau_h=R_\star/\varv_\text{SN}$ (where we use $\varv_\text{SN}$ as an averaged maximum 
velocity of the forward shock front at the specified time), and the effective light 
curve timescale $\tau_m=\sqrt{2\tau_0\tau_h}$ \citep{1982ApJ...253..785A}. 
The SN luminosity is then determined by
\begin{equation}
\label{lumin}
L_\text{SN}(t)=L_\text{SN}(0)\,\varphi(t)\sim \frac{f E_\text{SN}}{\tau_0}\,\varphi(t),
\end{equation}
where $f$ is numerical factor (ratio of the initial thermal energy $E_\text{th}(0)$  
to the total energy $E_\text{SN}$; we may relevantly choose $f=0.5$). 
The dimensionless function $\varphi(t)$ is \citep[][]{2012ApJ...746..121C,mcdowell18}
\begin{equation}
\label{lumin2}
\varphi(t)\approx\text{e}^{-(t/\tau_0+t^2/\tau_m^2)}\int_0^t \frac{\dot{Q}(t^\prime)}{L_\text{SN}(0)}\,\text{e}^{(t^\prime/\tau_0+{t^\prime}^2/\tau_m^2)}\left(\frac{\tau_h+2t^\prime}{\tau_m^2}\right) \d t^\prime.
\end{equation}

In the bottom panel of Fig.~\ref{fig:lc}, we show the resulting time evolution 
of $L_\text{SN}$. We see that taking into account diffusion converts $\dot{Q}$ 
into gradually rising light curves with bumps and wiggles smoothed out, although model 
C retains some of the small scale structure. However, we do not see any light curve 
feature that would allow us to distinguish between different CSM geometries and different 
radial density profiles of spherically symmetric CSM.

Finally, we emphasize that shock interaction in our models is not occurring over the 
full solid angle of the SN ejecta. As a result, the part of the SN ejecta not interacting 
with the CSM will radiate similarly to a normal SN. This could lead to two-component light 
curves, similarly to what was suggested for luminous red novae \citep{metzger17} or kilonovae 
\citep[e.g.,][]{kasen17}. To illustrate this point, we show in the bottom panel of 
Figure~\ref{fig:lc} theoretical light curve of a SN powered by the decay of $0.28\,\msun$ 
of radioactive nickel. This light curve was obtained by using a different form of $\dot{Q}$ 
in Equation~(\ref{lumin2}). We see that for models B1 and B2a, the CSM is positioned 
sufficiently far away from the progenitor so that the observed light curve would likely 
have a first recombination/radioactivity powered peak, followed by second peak powered by 
shock interaction. For the remaining models, the CSM is so close and so dense that the shock 
interaction dominates.

\subsection{Spectral line profiles}
\label{lineprofiles}
\begin{figure}
\centering
\includegraphics[width=0.45\textwidth]{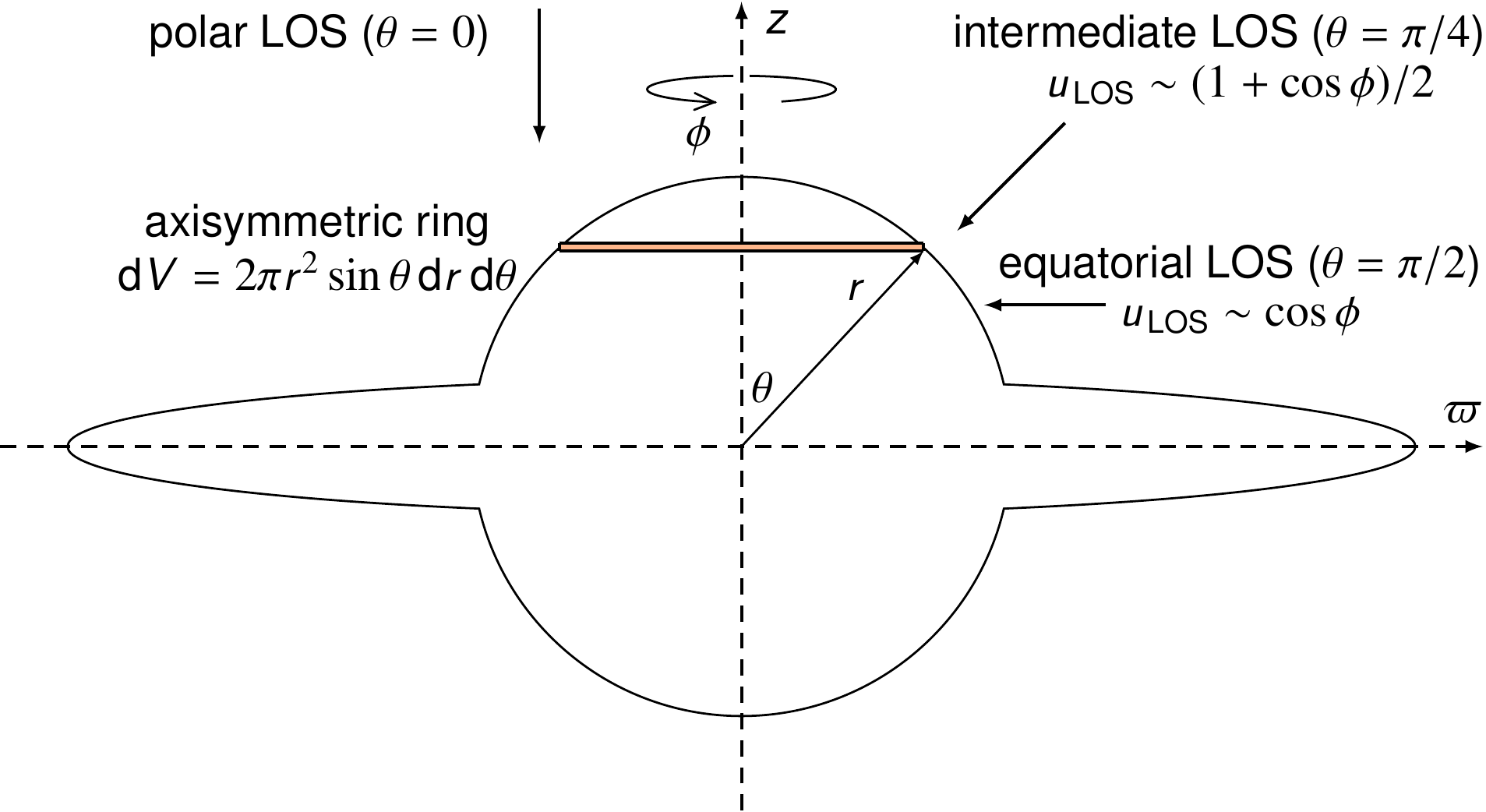}
\caption{Schematic picture of the directions when calculating line-of-sight velocity distributions.}
\label{losvelos} 
\end{figure}
Spectral line profiles can provide more insight into the ejecta geometry than 
disk-integrated light curves. However, calculating spectral line profiles in rapidly 
and differentially expanding medium out of local thermodynamic equilibrium is a complicated 
problem, which we do not attempt to solve here. Our goal is to provide guidance for how 
observed spectral line profiles relate to different CSM geometries. We are mostly 
interested in obtaining estimates of line profiles at late times, when the SN ejecta 
should be nearly transparent for radiation. 

We obtain approximate line profiles by calculating volume-weighted histograms of line-of-sight 
velocities $u_\text{los}$ for $\theta = 0$, $\pi/4$, and $\pi/2$. The geometry of the 
problem is illustrated in Fig.~\ref{losvelos}. Our approach follows the simple examples 
in \citet{jerkstrand17}. We neglect absorption within the ejecta and assume that the 
emissivity does not depend on local density and temperature. Furthermore, we excise 
dense inner parts of the SN envelope, which correspond to the helium core. We determined 
approximate helium core radii with SNEC to be  
$8.6\times 10^{14}\,\text{cm}\approx 12\,R_\star$ at $100$\,d,
$1.8\times 10^{15}\,\text{cm}\approx 25\,R_\star$ at $200$\,d, 
$2.6\times 10^{15}\,\text{cm}\approx 36\,R_\star$ at $300$\,d, 
and $3.3\times 10^{15}\,\text{cm}\approx 46\,R_\star$ at $400$\,d. 
Since our simulations are axisymmetric, we extend the dimensionality and add 
the azimuthal dependency by dividing each quadrant of the model to 24 azimuthal 
($\phi$-direction) intervals. More details of the calculation of $u_\text{LOS}$ 
are shown in Fig.~\ref{losvelos}. The resulting $u_\text{LOS}$ distributions are 
binned to approximately 360 -- 960 bins within the total velocity range of 
$\approx\pm 10^4\,\text{km}\,\text{s}^{-1}$.

In order to build understanding of the spectral line profiles and to test our method, 
we first calculated the velocity distributions for simple configurations such as 
a homogeneous expanding sphere with constant density and homologous radial velocity 
profile. This configuration has the expected parabolic shape \citep{jerkstrand17}. 
We then continued by adding artificial polar velocity components of different magnitudes. 
We also tested the more realistic case of spherically symmetric expanding SN without 
CSM with the input parameters corresponding to the progenitor parameters of our models. 
Finally, we address the issue of how to distinguish between the SN ejecta and the 
unshocked CSM. The unshocked CSM has typically much lower velocities than the SN ejecta 
and would contaminate only the bins near $u_\text{LOS} \approx 0$. We manually remove 
most of this effect, but caution should be taken when interpreting results near 
$u_\text{LOS} \approx 0$.

\begin{figure*}
\centering
\begin{tabular}{rc}
A & \raisebox{-.92\height}{\includegraphics[width=0.58\textwidth]
{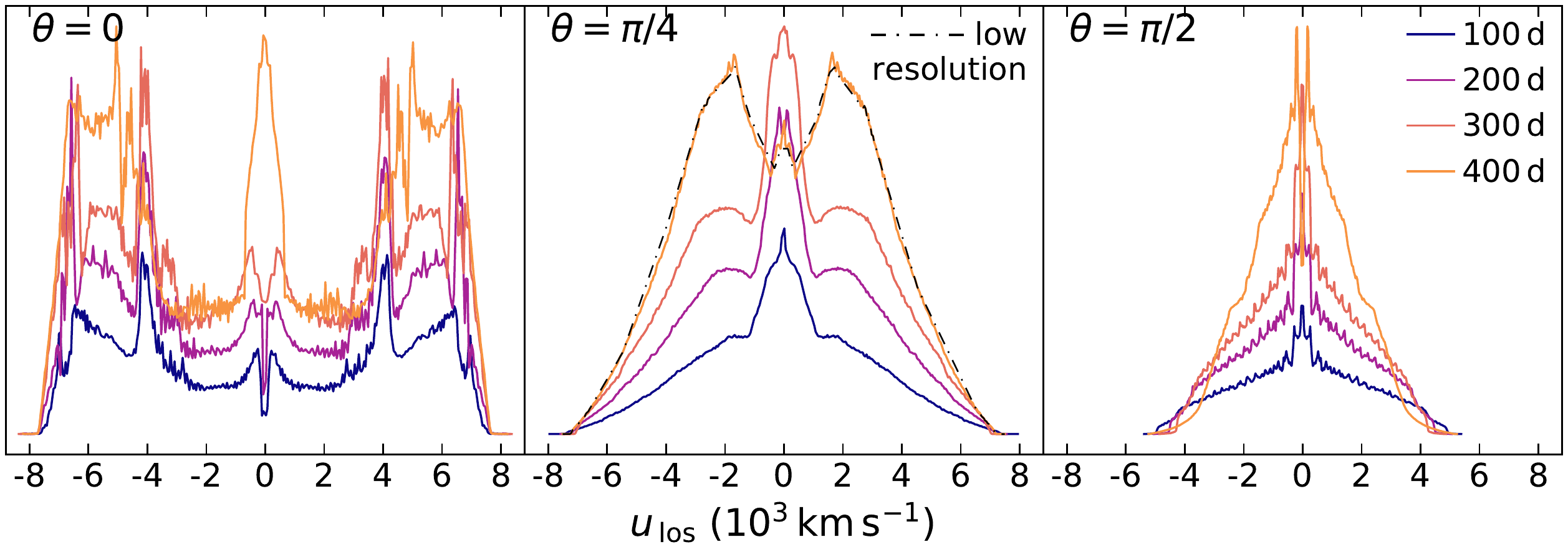}}\\
B1 & \raisebox{-.92\height}{\includegraphics[width=0.58\textwidth]
{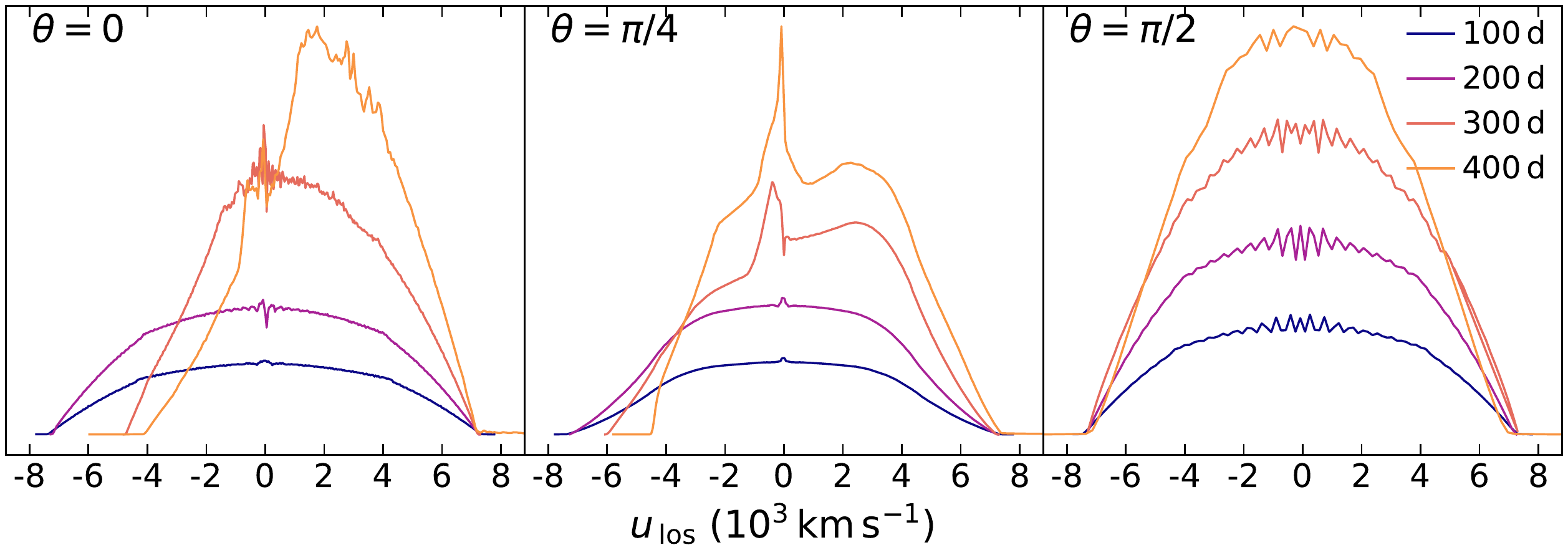}}\\
B2a & \raisebox{-.92\height}{\includegraphics[width=0.58\textwidth]
{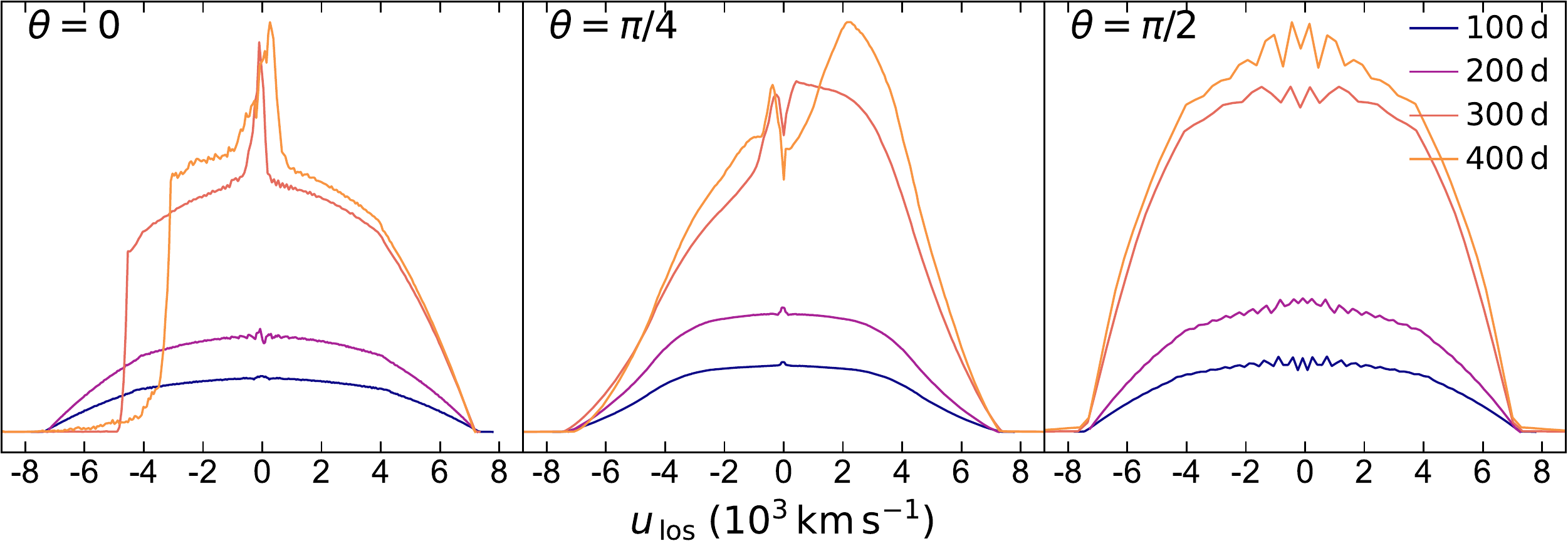}}\\
B2b & \raisebox{-.92\height}{\includegraphics[width=0.58\textwidth]
{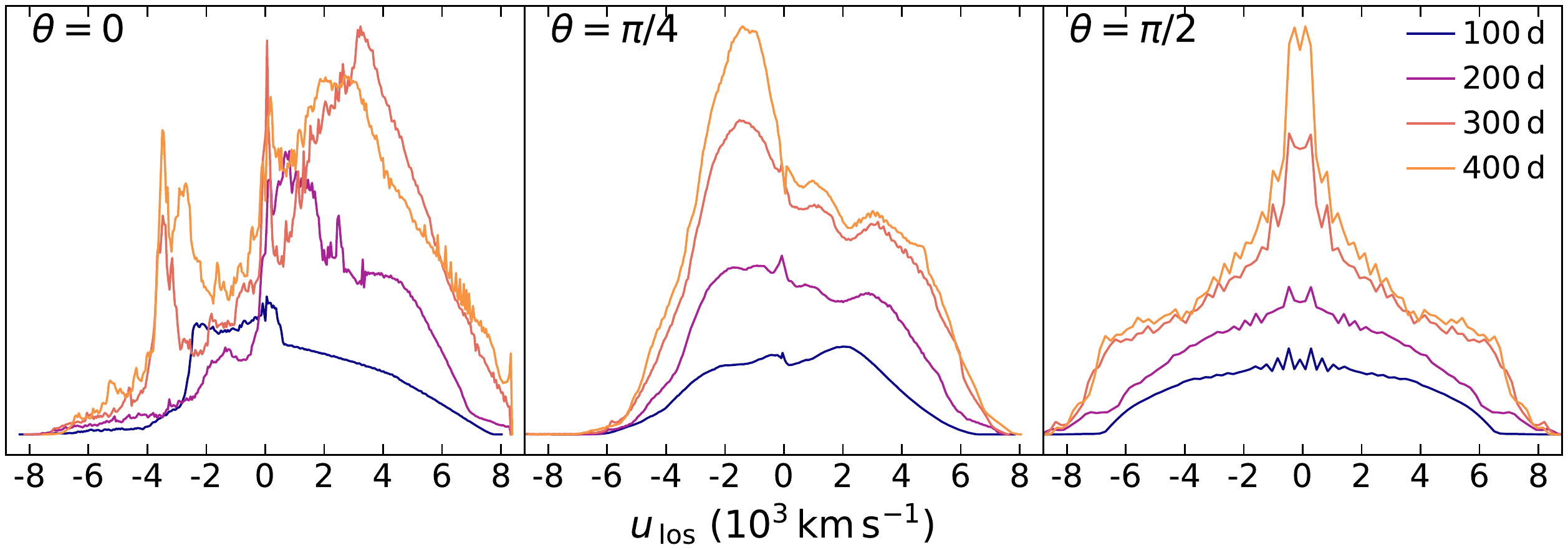}}\\
B3  & \raisebox{-.92\height}{\includegraphics[width=0.58\textwidth]
{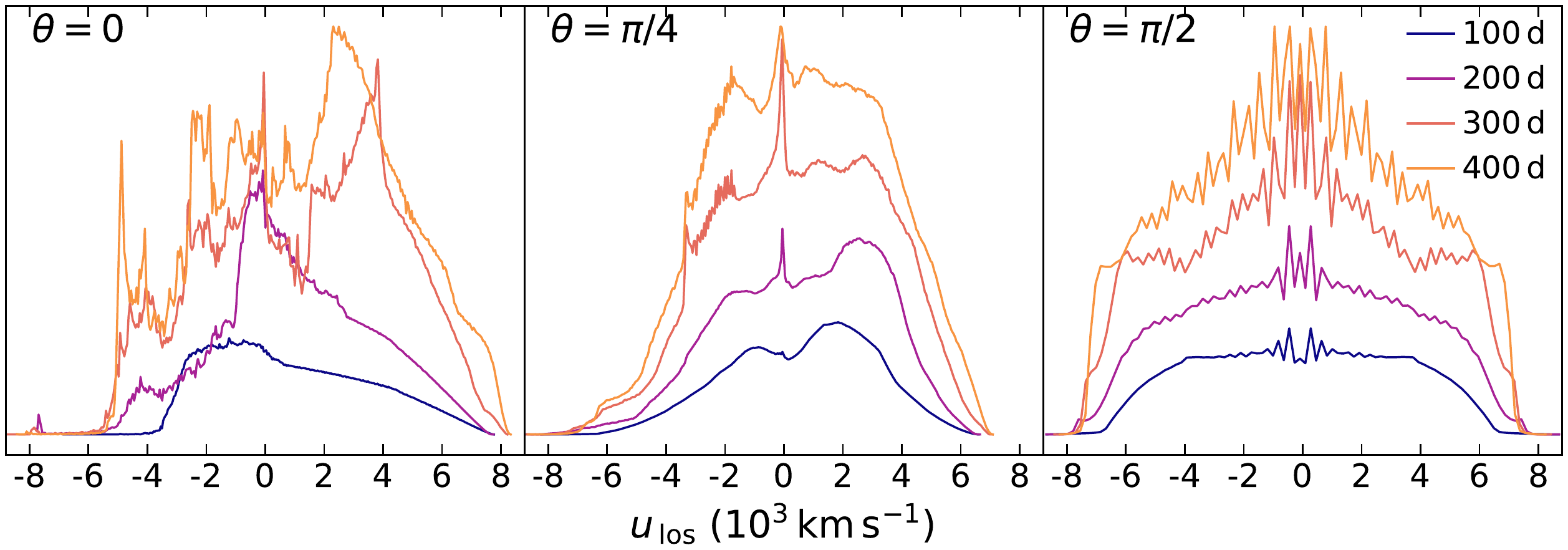}}\\
C   & \raisebox{-.92\height}{\includegraphics[width=0.58\textwidth]
{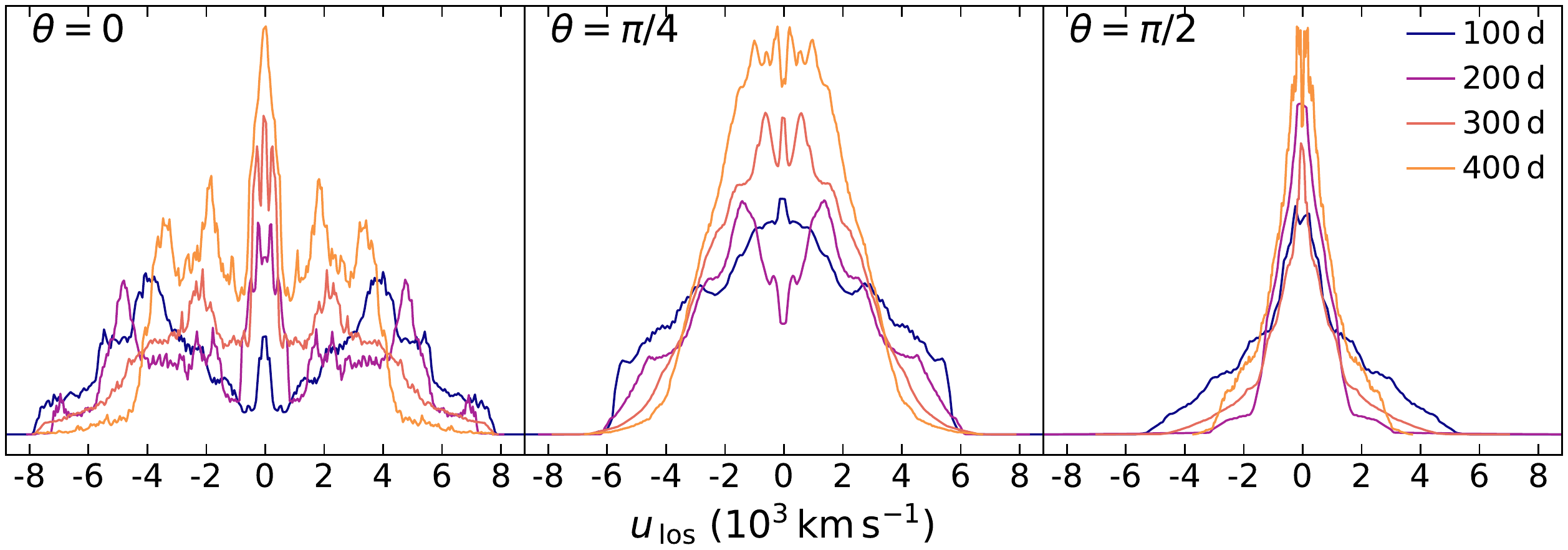}}\\
\end{tabular}
\caption{Line-of-sight velocity distributions for our models. Each row corresponds 
to a different model, labeled on the left, and each column represents different 
viewing angle $\theta$. Colored lines are results for different simulation times with 
legend given in the plot. The distributions are normalized to ensure clarity of the 
presentation and hence we do not display units on the vertical axes. The distributions 
are plotted on a linear scale. Some of the features around $|u_\text{LOS}| \approx 0$ 
are an artifact of subtraction of CSM that has not collided with the SN ejecta. 
The black dash-dotted line in the middle panel of model A illustrates the 400~d 
lower resolution model.}
\label{spectra} 
\end{figure*}

In Figures~\ref{spectra}, we show the calculated histograms of line-of-sight velocities 
for our models at a range of times and for three lines of sight, 
$\theta = 0$, $\pi/4$, and $\pi/2$. Velocity for even higher viewing angles can 
be obtained by mirroring the lines in Fig.~\ref{spectra} around zero, for example, 
for $\theta=3\pi/4$ the velocity distribution would look like for $\theta=\pi/4$ 
but with $u_\text{LOS} \rightarrow -u_\text{LOS}$. We note that at later times the 
fastest SN ejecta have already left our computational grid and therefore the histograms 
are effectively cut at certain value of $|u_\text{LOS}|$. The velocity profiles are 
always symmetric around $u_\text{LOS}$ for $\theta=\pi/2$, because the CSM distributions 
are rotationally symmetric around the $z$ axis. 

For model A (circumstellar disk) and $\theta=0$ 
(looking from the top side of Fig.~\ref{fig2Ddisk}), we see the expected pattern 
of two peaks located symmetrically at high positive and negative $u_\text{LOS}$. 
The double-peaked pattern is less expressed from $\theta=\pi/4$ and not visible 
from $\theta = \pi/2$, because the velocity asymmetry is not aligned with the line of sight. 

Models of the B series exhibit the greatest asymmetry between positive and 
negative $u_\text{LOS}$, because the colliding wind shell is positioned on one side 
of the progenitor. In model B1, after the interaction starts at $t\gtrsim 250$\,days, 
negative $u_\text{LOS}$ is suppressed because the material moving toward the observer 
is intercepted by the colliding wind shell. The shell is curved toward the progenitor 
and blocks ejecta over a wide range of solid angles and spatial scales. As a result, 
the suppression of negative $u_\text{LOS}$ is relatively smooth. In model B2a, the 
fastest ejecta with small momentum are completely deflected by the shell. As a result, 
there is a sharp drop in the distribution for negative $u_\text{LOS}$ and $\theta = 0$. 
For $\theta = \pi/4$, the distribution is smoother and similar to model B1. Model B2b 
has the shell positioned closer to the SN, which means more time for the development 
of hydrodynamical instabilities. As a result, the velocity distribution for 
$t\lesssim 250$\,days is smooth and similar to model B2a but becomes rougher as the 
instabilities develop at later times. The peak at $u_\text{LOS} < 0$ that gets relatively 
stronger over time arises from the ejecta that have broken through the shell and continue 
to expand toward the observer. Model B3 is qualitatively similar to B2b except that 
the ejecta that break through the standoff point of the shell are moving slower and 
therefore the peak at negative $u_\text{LOS}$ is weaker. Interestingly, even at late 
times, the distribution for $\theta=\pi/4$ remains nearly symmetric. To summarize, 
interactions with colliding wind shells lead to asymmetric multipeaked velocity 
distributions, where the relative strengths of the red and blue wings depend on the 
viewing angle and evolve in time. In some cases, the strongest peaks evolve from the 
blue to the red side or in the opposite way.

Finally, model C exhibits multiple peaks symmetrically positioned at positive and 
negative $u_\text{LOS}$, because the bipolar nebula has mirror symmetry with respect 
to the $z=0$ plane. Despite a very different CSM distribution, the $u_\text{LOS}$ 
distributions resemble the circumstellar disk in model A, albeit with a number of smaller  
peaks at intermediate velocities.

\subsection{Polarization signatures}
\label{polarization}
\begin{table*}
\caption{Values of the shape factor $\gamma$, averaged optical depth $\bar{\tau}$, and the 
polarization degree $P_R$ for the models at four times, 
$t= 100,\,200,\,300,\text{ and }400\,\text{d}$. $P_R$ is given for $\theta=\pi/2$, 
which gives maximum $P_R$.}
\label{table1}
\begin{center}
\bgroup
\setlength{\tabcolsep}{5.25pt}
\def\arraystretch{1.12}
\begin{tabular}{c|ccr|ccr|ccr|ccr}
\hline
\hline
Model & \multicolumn{3}{c|}{$t= 100\,\text{d}$}& \multicolumn{3}{c|}
{$t= 200\,\text{d}$}& \multicolumn{3}{c|}{$t= 300\,\text{d}$}& \multicolumn{3}{c}
{$t= 400\,\text{d}$}\\\cline{2-13}
   & $\gamma$  & $\bar{\tau}$ & $P_R$ (\%) 
   & $\gamma$  & $\bar{\tau}$ & $P_R$ (\%) 
   & $\gamma$  & $\bar{\tau}$ & $P_R$ (\%)
   & $\gamma$  & $\bar{\tau}$ & $P_R$ (\%)\\\hline
A & $0.3448$ & $0.1941$ & $-0.6677$ & $0.3457$ & $0.4047$ & $-1.5014$ & $0.3459$ 
& $0.5189$ & $-1.9563$ & $0.3468$ & $0.5192$ & $-2.1004$ \\
B1 & $0.3330$ & $0.1572$ & $0.0157$ & $0.3327$ & $0.3881$ & $0.0737$ & $0.3320$ 
& $0.4712$ & $0.1819$ & $0.3314$ &  $0.5066$ & $0.2989$ \\
B2a & $0.3327$ & $0.1602$ & $0.0295$ & $0.3326$ & $0.4032$ & $0.0848$ & $0.3314$ 
& $0.5103$ & $0.2914$ & $0.3308$ & $0.5177$ & $0.3876$ \\
B2b & $0.3292$ & $0.1530$ & $0.1897$ & $0.3281$ & $0.2365$ & $0.3713$ & $0.3285$ 
& $0.3310$ & $0.4753$ & $0.3289$ & $0.4084$ & $0.5459$\\
B3 & $0.3423$ & $0.1446$ & $-0.3890$ & $0.3430$ & $0.2062$ & $-0.5980$ & $0.3395$ 
& $0.3258$ & $-0.6027$ & $0.3386$ & $0.4031$ & $-0.6369$ \\
C & $0.3532$ & $0.1810$ & $-1.0788$ & $0.3447$ & $0.3899$ & $-1.3330$ & $0.3435$ 
& $0.5017$ & $-1.5238$ & $0.3462$ & $0.4888$ & $-1.8868$ \\
\hline
\end{tabular}
\egroup
\end{center}
\end{table*} 

\begin{figure}
\centering
\includegraphics[width=0.4\textwidth]{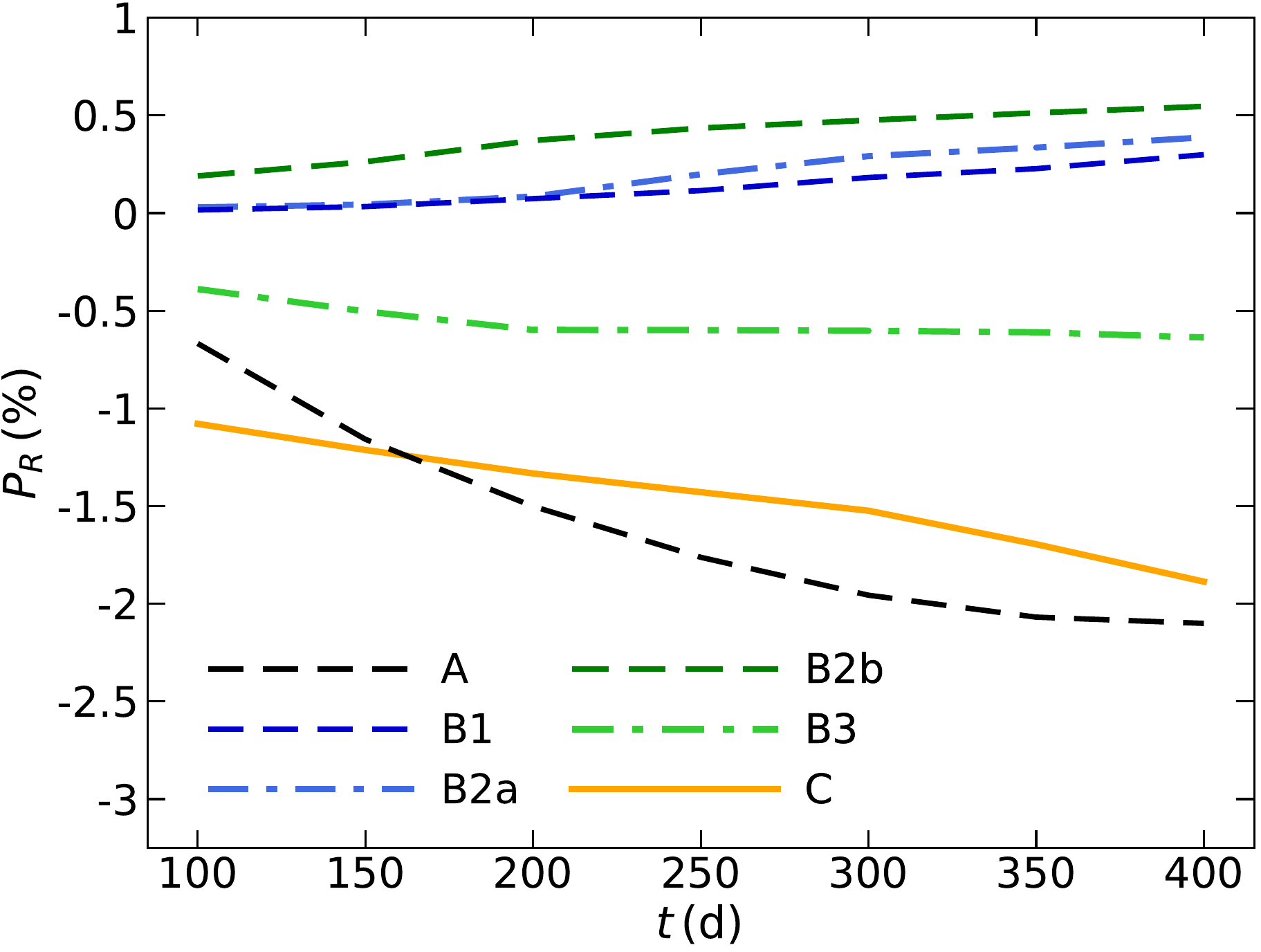}
\caption{Evolution of relative polarization
degree for our models. Values at selected times are given in Table~\ref{table1}.}
\label{polar_fig} 
\end{figure}
We estimate the polarization degree $P_R$ of the SN ejecta using the analytical 
prescriptions of \citet{1977A&A....57..141B} and
\citet{1978A&A....68..415B}, which were derived under the assumption of Thomson scattering 
in an optically thin envelope irradiated by a central source. These approximations 
are certainly very crude for SN ejecta at late times, but allow us to get at least a 
relative assessment of the asphericity induced by the  different CSM geometries. 
The polarization degree is given by
\begin{align}
\label{polar1}
P_R\simeq\bar{\tau}(1-3\gamma)\sin^2 \theta,
\end{align}
where $\theta$ is the inclination with the convention used in this paper, namely 
$\theta=0$ when viewed pole-on and $\theta=\pi/2$ when viewed equator-on. The averaged 
Thomson scattering optical depth $\bar{\tau}$ of the envelope and  the shape 
factor $\gamma$ are
\begin{align}\label{polar2}
\bar{\tau}=\frac{3}{16}\sigma_T
\int_{R_{\text{He}}}^{\infty}
\int_{\mu=-1}^{1} n\,\text{d}r\,\text{d}\mu,\quad
\gamma=\frac{\int_{R_{\text{He}}}^{\infty}\int_{\mu=-1}^{1}n\mu^2\,\text{d}r\,\text{d}\mu}
{\int_{R_{\text{He}}}^{\infty}
 \int_{\mu=-1}^{1}n\,\text{d}r\,\text{d}\mu},
\end{align}
where $\sigma_T$ is the Thomson scattering cross-section, $n(r,\mu)$ is the electron 
number density, and $\mu=\cos\,\theta$. Assuming complete ionization of a pure hydrogen 
envelope, the electron number density is  $n(r,\mu)=\rho(r,\mu)/m_\text{H}$, 
where $m_\text{H}$ is the mass of hydrogen atom. Equation~\eqref{polar2} implies 
$\gamma =1/3$ for any spherically symmetric distribution, while $\gamma<1/3$ ($P_R>0$) 
in case of oblate, and $\gamma>1/3$ ($P_R<0$) in case of prolate mass distributions 
\citep{1977A&A....57..141B,1978A&A....68..415B}. We eliminate the helium-dominated 
core from the calculation to emphasize the asphericity of the envelope and we denote 
the radius of the helium core as the  lower limit of radial integration, $R_{\text{He}}$, 
in Eq.~(\ref{polar2}). We determined $R_\text{He}$ in the same way as in 
Sect.~\ref{lineprofiles}.

Our results are shown in Fig.~\ref{polar_fig} and listed for four selected times, 
$t\approx 100$, $200$, $300$, and $400$\,days, in Table~\ref{table1}. 
First, we look at models B1 and B2a at $100$\,days, where our numerically calculated 
$P_R$ is $\approx  0.03\%$. We would expect the bulk of the ejecta to be spherically 
symmetric with $P_R=0$ at this time because the SN ejecta have not reached the CSM 
at this epoch. Consequently, $P_R \approx 0.03\%$ roughly corresponds to the numerical 
noise of our polarization estimates. Looking now at the results, we see that models 
A and C have negative $P_R$ reaching about $2\%$ at the end of our simulations. 
Polarization implies prolate distributions, which is caused by deflection of SN ejecta 
to polar regions by circumstellar disk (model A) or dense equatorial waist (model C).  
Models B1, B2a, and B2b exhibit positive $P_R$, which remains low, $P_R \lesssim 0.5\%$, 
at late times. Naturally, these models have oblate mass distribution because the 
colliding wind shell deflects material to the equatorial plane. Finally, model B3 
reaches $P_R \approx -0.6\%$, which suggests prolate mass distribution likely caused 
by SN ejecta interaction near the standoff point. 

\subsection{Comparison with observed supernovae}
\label{compare}

The disk-integrated quantities estimated from our simulations broadly agree with what is 
observed: our approximate light curves in Fig.~\ref{fig:lc} peak around 
$10^{42}$--$10^{43}$\,ergs\,s$^{-1}$ and evolve slowly similarly to 
some observed overplotted in the same figure. The peak luminosity and the slope 
of the theoretical light curves depend on the CSM properties, which we do not systematically 
vary in this work. The actual light curve could also be a combination of radioactively-powered 
component (schematically indicated by the thin solid black line in the bottom panel of 
Fig.~\ref{fig:lc}) and the shock interaction component calculated in one of our models. 
The polarizations in Table~\ref{table1} match the continuum polarizations of $1$--$3\%$ 
observed in interacting SNe \citep[e.g.,][]{leonard00,bilinski18}. However, these quantities 
depend on unconstrained parameters like density profile or inclination and therefore 
do not provide a clear-cut signature differentiating between CSM geometries. Consequently, 
we focus on the line-of-sight velocity distributions as a proxy for late-time spectral 
line profiles. There are a number of interacting SNe with late-time spectral line 
profiles implicating aspherical CSM. Few recent examples include SN2007od 
\citep{andrews10,inserra11}, SN2012ab \citep{bilinski18}, SN2013L \citep{andrews17,taddia20}, 
SN2013ej \citep{bose15,huang15,yuan16,mauerhan17}, SN2014G \citep{terreran16}, 
SN2014ab \citep{bilinski20}, iPTF14hls \citep{andrews18},  KISS15s \citep{kokubo19},  
SN2017eaw \citep{szalai19,weil20}, and SN2017gmr \citep{andrews19}. Typically, 
double-peaked line profiles are attributed to disk- or torus-like geometry 
\citep[e.g.,][]{gerardy00,jerkstrand17}. Double-peaked profile is also prominently 
seen in our model A in Fig.~\ref{spectra}. A boxy, flat-topped profile with a 
possibility of double-peaked horns was argued to arise from bipolar lobes similar 
to what is seen in $\eta$~Car. An example is an eruption of a SN impostor UGC 2773-OUT 
\citep{smith16}. Our model C in Fig.~\ref{spectra} remotely resembles such profiles, 
but with relatively strong dependence on the viewing angle.

Suppression of the red side of a nebular spectral line is often attributed to dust formation, 
which more effectively blocks the light coming from the more distant receding side of 
the SN \citep[e.g.,][]{lucy89,sugerman06,smith08}. However, in some cases the observations 
indicate that there is a genuine asymmetry in the SN, or more specifically, the SN 
lacks either rotational symmetry with respect to an axis or mirror symmetry with 
respect to an equatorial plane. One example is SN2013L, where \citet{andrews17} 
observed the same blueshifted asymmetric line profile in H$\alpha$ and Pa$\beta$ 
that cannot be explained by internal wavelength-dependent dust obscuration. 
Even more striking example is PTF11iqb analyzed by \citet{smith15}. In the first few 
days after the explosion, PTF11iqb showed Wolf--Rayet-like spectral features likely 
arising from flash-ionized CSM of the inner wind of its red supergiant progenitor. 
For the next $\sim 100$\,days, the spectrum and light curve resembled a normal 
Type II-P SN. In the nebular phase, H$\alpha$ emission initially showed a blueshifted peak, 
but after $\sim 500$\,days a redshifted peak appeared and eventually dominated H$\alpha$ 
emission. In addition, H$\alpha$ emission exhibited several smaller bumps and peaks. 
\citet{smith15} could not explain the late evolution by disappearance of the dust 
and instead argued for a disk- or torus-like CSM with enhanced density on the 
more distant redshifted side of the SN. 

Instead of an azimuthally-asymmetric disk, there are several reasons why an 
interaction with a colliding wind shell could explain PTF11iqb. First, the colliding 
wind shell occurs only on one side of the SN and naturally satisfies the condition 
of equatorial plane asymmetry. Second, the Wolf--Rayet-like signatures could arise 
either in the dense slow wind of the progenitor or in the density enhancement in the 
colliding wind shell, similarly to what \citet{kochanek19} suggested for SN2013fs. 
Third, the standoff point of the colliding wind shell might be located relatively 
close to the SN, which means that the CSM interaction begins early with a sharp peak 
and continues to gradually decrease over time (models B2b and B3 in Fig.~\ref{fig:lc}). 
Fourth, since the shock interaction occurs only over a small fraction of the solid angle, 
the shock can be embedded in the ejecta to hide the narrow lines and produce a relatively 
normal-looking plateau, similarly to what is argued for a disk-like CSM. Finally, 
the evolution of blue and red peaks of nebular H$\alpha$ resembles models B2b and B3 
viewed either from $\theta=0$ or $\theta=3\pi/4$ (equivalent to $\theta=\pi/4$ 
with velocity distribution flipped around the origin). 

There are other events, where colliding wind shell might be a better explanation 
for the observations than circumstellar disk. Following the reasoning of \citet{smith15}, 
SN1998S might be explained by a similar geometry as PTF11iqb, but viewed from a different 
viewing angle. Recently, \citet{bilinski20} presented observations of SN2014ab, which 
show nearly identical H$\alpha$ and Pa$\beta$ profiles with strong blueshifted component, 
implying a lack of symmetry between the near and far side of the SN. The event shows 
little polarization suggesting circular symmetry from our line of sight. \citet{bilinski20} 
argued that the lack of polarization is due to our viewing angle near the axis of symmetry 
(for example, looking from $\theta=0$ at our model A) and the spectral line asymmetry 
is due to internal absorption in the shock interaction region, which hides its farther 
receding side. Interestingly, our colliding wind models consistently predict asymmetric 
velocity distributions from most viewing angles and noticeably smaller polarization 
degrees than circumstellar disk or bipolar lobes models. Although the observed 
luminosities of PTF11iqb and SN2014ab are somewhat smaller than what we predict 
in our models (bottom panel of Fig.~\ref{fig:lc}), we expect that better agreement 
could be reached by varying some of the parameters of our models. For example, 
lowering $\alpha$ in Eq.~(\ref{bow4}) would decrease the surface density and total 
mass of the colliding wind shell while maintaining the shape, which would lead to 
lower luminosity from the shock and better agreement with the observed light curves.
 
\section{Conclusions}
\label{conclude} 

We have performed two-dimensional axisymmetric hydrodynamic simulations of 
spherically symmetric SN ejecta colliding with aspherical CSM. The main improvements 
over the previous works are realistic initial density and velocity profiles of 
the SN ejecta and a wider range of CSM geometries. In particular, for the first time 
we studied shock interaction with a colliding wind shell in a binary star and compared 
the results to SN interactions with  circumstellar disk and bipolar lobes. The 
typical CSM masses on our computational grid are $10^{-3}$ to $10^{-2}\,\msun$.  
Snapshots of density, radial and polar velocity, temperature, and shock heating 
are summarized in Figs.~\ref{fig2Ddisk} - \ref{lobescolortiles} and in 
Sect.~\ref{overhydro}.  All our models exhibit deceleration of the expanding 
SN ejecta by the CSM and the ensuing deflection of the explosion to the directions 
of the least resistance. The hydrodynamics involves oblique and deflected shocks 
and their clustering, shearing motions accompanied by Kelvin--Helmholtz instabilities, 
and shock propagation through density gradients leading to Rayleigh--Taylor-like 
instabilities. We saw that the expanding material often wraps around the denser parts 
of the CSM.

Based on our hydrodynamical simulations, we estimated three observables of CSM interaction
in SNe: the shock power and the related bolometric luminosity, distribution of 
line-of-sight velocities as a proxy for late-time spectral line profiles, and the 
degree of polarization. The shock energy deposition closely traces the course of the 
interaction and reaches up to $10^{44}$\,ergs\,s$^{-1}$. If embedded inside an 
optically thick envelope, the energy generated by the shock gradually diffuses out of 
the SN ejecta, which we estimate using an analytic one-zone model. The resulting 
bolometric luminosities are in the range of $10^{42}$ to $10^{43}$\,ergs\,s$^{-1}$ 
(Fig.~\ref{fig:lc} and Sect.~\ref{lightcurves}), which is in the range of what is observed 
in Type IIn SNe (Fig.~\ref{Nyholm}). The time dependence of shock power shows short-term 
fluctuations or peaks with amplitudes $\lesssim 10\%$, which get smoothed and erased 
if the shocks are embedded in and reprocessed by the SN envelope. Interaction with 
bipolar lobes, where the CSM is structured in several concentric shells, leads to 
more prominent fluctuations. Our models thus do not readily explain the bumps and 
wiggles observed in some interacting SNe (Fig.~\ref{Nyholm}). Perhaps coupling 
radiation to the hydrodynamics would allow for easier escape of the shock-generated 
radiation through the crevices created by the hydrodynamic instabilities. Alternatively, 
the instabilities could be amplified by more realistic initial conditions taking into 
account the clumpiness of the CSM \citep[e.g.,][]{calderon16,calderon20}.

The distribution of line-of-sight velocities (Fig.~\ref{spectra} and Sect.~\ref{lineprofiles}) 
has the greatest discriminating power between different CSM geometries studied here. 
Our models show the expected double-peaked profile for the circumstellar disk and 
symmetric multipeaked flat-top profile for the bipolar lobes. The colliding wind 
shell is positioned only on one side of the SN and could naturally explain blue--red 
asymmetry of late-time line profiles, which cannot be readily explained by internal 
obscuration due to dust. An example of such object is PTF11iqb \cite[Sect.~\ref{compare} 
and][]{smith15}.  The small solid angle subtended by the interaction regions could 
lead to engulfment of the shock by the SN ejecta, which might hide the narrow lines 
and make the shock essentially an internal power source inside the envelope.

Our estimates of the degree of polarization (Tab.~\ref{table1} and Sect.~\ref{polarization}) 
give values similar to what is observed \citep[e.g.,][]{2011MNRAS.415.3497D,2019ARA&A..57..305G}.
CSM in the form of circumstellar disk and bipolar lobes leads to prolate shape of the ejecta 
and maximum polarization on the level of $1$--$2\%$. Interaction with a colliding wind shell 
leads to smaller amounts of polarization $\lesssim 0.5\%$ and usually oblate shapes. 
Despite these differences, the estimates of polarization degree of our models are not 
sufficiently different from each other to discriminate between different CSM geometries 
on their own, especially when taking into account unconstrained degrees of freedom such 
as the viewing angle and parameters of the CSM density distributions.

To summarize, we performed hydrodynamic-only simulations to explore and widen the range 
of CSM geometries considered for interacting SNe. We recovered expected results 
for circumstellar disk and bipolar lobes. Our results suggest that colliding wind 
shells are particularly promising for explaining more complicated asymmetries and time 
evolution observed in some SNe. Occurrence rates of colliding wind shells around SN 
progenitors should be estimated, for example, based on the binary population synthesis 
models \citep[e.g.,][]{eldridge08}. More sophisticated simulations including proper treatment 
of radiation should be performed to provide more realistic predictions for observations. 
With sufficiently developed theory to robustly detect and characterize colliding wind 
shells interacting with SN explosions, it might be possible to characterize binary 
companions to SNe in a new way. For example, modeling of the observed light curves 
and spectral line profiles might provide the time when the shock reaches the standoff 
point of the colliding wind shell, which is proportional to the physical distance from 
the progenitor, as well as  the orientation of the shell. With these estimates, we could 
infer the separation between the two binary components and the wind momentum of the companion, 
which would constrain its evolutionary state.

\begin{acknowledgements}
We thank to Dr. Milan Prvák for technical support and useful improvements in computational 
process and in preparation of the draft of this paper. PK and OP were supported by the 
grant Primus/SCI/17. OP was additionally supported by Horizon 2020 ERC Starting Grant 
``Cat-In-hAT'' (grant agreement \#803158) and INTER-EXCELLENCE grant LTAUSA18093 from 
the Ministry of Education, Youth, and Sports. This work was also supported by the 
grant GA\v{C}R 18-05665S. Computational resources were supplied by the project 
"e-Infrastruktura CZ"
(e-INFRA LM2018140) provided within the program Projects of Large Research,
Development and Innovations Infrastructures.
\end{acknowledgements}

\bibliographystyle{aa} 
\bibliography{bibliography} 

\begin{thebibliography}{109}
\expandafter\ifx\csname natexlab\endcsname\relax\def\natexlab#1{#1}\fi

\bibitem[{{Anderson}(2019)}]{anderson19}
{Anderson}, J.~P. 2019, \aap, 628, A7

\bibitem[{{Andrews} {et~al.}(2010){Andrews}, {Gallagher}, {Clayton},
  {Sugerman}, {Chatelain}, {Clem}, {Welch}, {Barlow}, {Ercolano}, {Fabbri},
  {Wesson}, \& {Meixner}}]{andrews10}
{Andrews}, J.~E., {Gallagher}, J.~S., {Clayton}, G.~C., {et~al.} 2010, \apj,
  715, 541

\bibitem[{{Andrews} {et~al.}(2019){Andrews}, {Sand}, {Valenti}, {Smith},
  {Dastidar}, {Sahu}, {Misra}, {Singh}, {Hiramatsu}, {Brown}, {Hosseinzadeh},
  {Wyatt}, {Vinko}, {Anupama}, {Arcavi}, {Ashall}, {Benetti}, {Berton},
  {Bostroem}, {Bulla}, {Burke}, {Chen}, {Chomiuk}, {Cikota}, {Congiu}, {Cseh},
  {Davis}, {Elias-Rosa}, {Faran}, {Fraser}, {Galbany}, {Gall}, {Gal-Yam},
  {Gangopadhyay}, {Gromadzki}, {Haislip}, {Howell}, {Hsiao}, {Inserra},
  {Kankare}, {Kuncarayakti}, {Kouprianov}, {Kumar}, {Li}, {Lin}, {Maguire},
  {Mazzali}, {McCully}, {Milne}, {Mo}, {Morrell}, {Nicholl}, {Ochner},
  {Olivares}, {Pastorello}, {Patat}, {Phillips}, {Pignata}, {Prentice},
  {Reguitti}, {Reichart}, {Rodr{\'\i}guez}, {Rui}, {Sanwal}, {S{\'a}rneczky},
  {Shahband eh}, {Singh}, {Smartt}, {Strader}, {Stritzinger}, {Szak{\'a}ts},
  {Tartaglia}, {Wang}, {Wang}, {Wang}, {Wheeler}, {Xiang}, {Yaron}, {Young}, \&
  {Zhang}}]{andrews19}
{Andrews}, J.~E., {Sand}, D.~J., {Valenti}, S., {et~al.} 2019, \apj, 885, 43

\bibitem[{{Andrews} \& {Smith}(2018)}]{andrews18}
{Andrews}, J.~E. \& {Smith}, N. 2018, \mnras, 477, 74

\bibitem[{{Andrews} {et~al.}(2017){Andrews}, {Smith}, {McCully}, {Fox},
  {Valenti}, \& {Howell}}]{andrews17}
{Andrews}, J.~E., {Smith}, N., {McCully}, C., {et~al.} 2017, \mnras, 471, 4047

\bibitem[{{Arcavi} {et~al.}(2017){Arcavi}, {Howell}, {Kasen}, {Bildsten},
  {Hosseinzadeh}, {McCully}, {Wong}, {Katz}, {Gal-Yam}, {Sollerman}, {Taddia},
  {Leloudas}, {Fremling}, {Nugent}, {Horesh}, {Mooley}, {Rumsey}, {Cenko},
  {Graham}, {Perley}, {Nakar}, {Shaviv}, {Bromberg}, {Shen}, {Ofek}, {Cao},
  {Wang}, {Huang}, {Rui}, {Zhang}, {Li}, {Li}, {Zhang}, {Valenti}, {Guevel},
  {Shappee}, {Kochanek}, {Holoien}, {Filippenko}, {Fender}, {Nyholm}, {Yaron},
  {Kasliwal}, {Sullivan}, {Blagorodnova}, {Walters}, {Lunnan}, {Khazov},
  {Andreoni}, {Laher}, {Konidaris}, {Wozniak}, \& {Bue}}]{arcavi17}
{Arcavi}, I., {Howell}, D.~A., {Kasen}, D., {et~al.} 2017, \nat, 551, 210

\bibitem[{{Aretxaga} {et~al.}(1999){Aretxaga}, {Benetti}, {Terlevich},
  {Fabian}, {Cappellaro}, {Turatto}, \& {della Valle}}]{aretxaga99}
{Aretxaga}, I., {Benetti}, S., {Terlevich}, R.~J., {et~al.} 1999, \mnras, 309,
  343

\bibitem[{{Arnett}(1980)}]{1980ApJ...237..541A}
{Arnett}, W.~D. 1980, \apj, 237, 541

\bibitem[{{Arnett}(1982)}]{1982ApJ...253..785A}
{Arnett}, W.~D. 1982, \apj, 253, 785

\bibitem[{{Bilinski} {et~al.}(2020){Bilinski}, {Smith}, {Williams}, {Smith},
  {Andrews}, {Clubb}, {Zheng}, {Filippenko}, {Fox}, {Hosseinzadeh}, {Howell},
  {Kelly}, {Milne}, {Sand}, {Hoffman}, {Leonard}, {Cargill}, {Casper},
  {Halevy}, {Kim}, {Kumar}, {Pina}, \& {Yuk}}]{bilinski20}
{Bilinski}, C., {Smith}, N., {Williams}, G.~G., {et~al.} 2020, arXiv e-prints,
  arXiv:2007.12134

\bibitem[{{Bilinski} {et~al.}(2018){Bilinski}, {Smith}, {Williams}, {Smith},
  {Zheng}, {Graham}, {Mauerhan}, {Andrews}, {Filippenko}, {Akerlof},
  {Chatzopoulos}, {Hoffman}, {Huk}, {Leonard}, {Marion}, {Milne}, {Quimby},
  {Silverman}, {Vink{\'o}}, {Wheeler}, \& {Yuan}}]{bilinski18}
{Bilinski}, C., {Smith}, N., {Williams}, G.~G., {et~al.} 2018, \mnras, 475,
  1104

\bibitem[{{Blondin} {et~al.}(1996){Blondin}, {Lundqvist}, \&
  {Chevalier}}]{blondin96}
{Blondin}, J.~M., {Lundqvist}, P., \& {Chevalier}, R.~A. 1996, \apj, 472, 257

\bibitem[{{Bose} {et~al.}(2015){Bose}, {Sutaria}, {Kumar}, {Duggal}, {Misra},
  {Brown}, {Singh}, {Dwarkadas}, {York}, {Chakraborti}, {Chandola},
  {Dahlstrom}, {Ray}, \& {Safonova}}]{bose15}
{Bose}, S., {Sutaria}, F., {Kumar}, B., {et~al.} 2015, \apj, 806, 160

\bibitem[{{Braun} {et~al.}(2012){Braun}, {Baade}, {Reimers}, \&
  {Hagen}}]{braun12}
{Braun}, K., {Baade}, R., {Reimers}, D., \& {Hagen}, H.~J. 2012, \aap, 546, A3

\bibitem[{{Brown} \& {McLean}(1977)}]{1977A&A....57..141B}
{Brown}, J.~C. \& {McLean}, I.~S. 1977, \aap, 57, 141

\bibitem[{{Brown} {et~al.}(1978){Brown}, {McLean}, \&
  {Emslie}}]{1978A&A....68..415B}
{Brown}, J.~C., {McLean}, I.~S., \& {Emslie}, A.~G. 1978, \aap, 68, 415

\bibitem[{{Calder{\'o}n} {et~al.}(2016){Calder{\'o}n}, {Ballone}, {Cuadra},
  {Schartmann}, {Burkert}, \& {Gillessen}}]{calderon16}
{Calder{\'o}n}, D., {Ballone}, A., {Cuadra}, J., {et~al.} 2016, \mnras, 455,
  4388

\bibitem[{{Calder{\'o}n} {et~al.}(2020){Calder{\'o}n}, {Cuadra}, {Schartmann},
  {Burkert}, {Prieto}, \& {Russell}}]{calderon20}
{Calder{\'o}n}, D., {Cuadra}, J., {Schartmann}, M., {et~al.} 2020, \mnras, 493,
  447

\bibitem[{{Chatzopoulos} {et~al.}(2012){Chatzopoulos}, {Wheeler}, \&
  {Vinko}}]{2012ApJ...746..121C}
{Chatzopoulos}, E., {Wheeler}, J.~C., \& {Vinko}, J. 2012, \apj, 746, 121

\bibitem[{{Chevalier}(1982)}]{chevalier82}
{Chevalier}, R.~A. 1982, \apj, 259, 302

\bibitem[{{Chevalier} \& {Soker}(1989)}]{chevalier89}
{Chevalier}, R.~A. \& {Soker}, N. 1989, \apj, 341, 867

\bibitem[{{Chornock} {et~al.}(2010){Chornock}, {Filippenko}, {Li}, \&
  {Silverman}}]{chornock10}
{Chornock}, R., {Filippenko}, A.~V., {Li}, W., \& {Silverman}, J.~M. 2010,
  \apj, 713, 1363

\bibitem[{{Chugai} \& {Danziger}(1994)}]{chugai94}
{Chugai}, N.~N. \& {Danziger}, I.~J. 1994, \mnras, 268, 173

\bibitem[{{Dessart} {et~al.}(2015){Dessart}, {Audit}, \& {Hillier}}]{dessart15}
{Dessart}, L., {Audit}, E., \& {Hillier}, D.~J. 2015, \mnras, 449, 4304

\bibitem[{{Dessart} \& {Hillier}(2011)}]{2011MNRAS.415.3497D}
{Dessart}, L. \& {Hillier}, D.~J. 2011, \mnras, 415, 3497

\bibitem[{{Dessart} {et~al.}(2016){Dessart}, {Hillier}, {Audit}, {Livne}, \&
  {Waldman}}]{dessart16}
{Dessart}, L., {Hillier}, D.~J., {Audit}, E., {Livne}, E., \& {Waldman}, R.
  2016, \mnras, 458, 2094

\bibitem[{{Eldridge} {et~al.}(2008){Eldridge}, {Izzard}, \&
  {Tout}}]{eldridge08}
{Eldridge}, J.~J., {Izzard}, R.~G., \& {Tout}, C.~A. 2008, \mnras, 384, 1109

\bibitem[{{Fransson} {et~al.}(2002){Fransson}, {Chevalier}, {Filippenko},
  {Leibundgut}, {Barth}, {Fesen}, {Kirshner}, {Leonard}, {Li}, {Lundqvist},
  {Sollerman}, \& {Van Dyk}}]{fransson02}
{Fransson}, C., {Chevalier}, R.~A., {Filippenko}, A.~V., {et~al.} 2002, \apj,
  572, 350

\bibitem[{{Gal-Yam}(2012)}]{2012Sci...337..927G}
{Gal-Yam}, A. 2012, Science, 337, 927

\bibitem[{{Gal-Yam}(2019)}]{2019ARA&A..57..305G}
{Gal-Yam}, A. 2019, \araa, 57, 305

\bibitem[{{Gayley} {et~al.}(1997){Gayley}, {Owocki}, \& {Cranmer}}]{gayley97}
{Gayley}, K.~G., {Owocki}, S.~P., \& {Cranmer}, S.~R. 1997, \apj, 475, 786

\bibitem[{{Gerardy} {et~al.}(2000){Gerardy}, {Fesen}, {H{\"o}flich}, \&
  {Wheeler}}]{gerardy00}
{Gerardy}, C.~L., {Fesen}, R.~A., {H{\"o}flich}, P., \& {Wheeler}, J.~C. 2000,
  \aj, 119, 2968

\bibitem[{{Goldman} {et~al.}(2017){Goldman}, {van Loon}, {Zijlstra}, {Green},
  {Wood}, {Nanni}, {Imai}, {Whitelock}, {Matsuura}, {Groenewegen}, \&
  {G{\'o}mez}}]{goldman17}
{Goldman}, S.~R., {van Loon}, J.~T., {Zijlstra}, A.~A., {et~al.} 2017, \mnras,
  465, 403

\bibitem[{{Gonz{\'a}lez} {et~al.}(2010){Gonz{\'a}lez}, {Villa}, {G{\'o}mez},
  {de Gouveia Dal Pino}, {Raga}, {Cant{\'o}}, {Vel{\'a}zquez}, \& {de La
  Fuente}}]{2010MNRAS.402.1141G}
{Gonz{\'a}lez}, R.~F., {Villa}, A.~M., {G{\'o}mez}, G.~C., {et~al.} 2010,
  \mnras, 402, 1141

\bibitem[{{Guillochon} {et~al.}(2017){Guillochon}, {Parrent}, {Kelley}, \&
  {Margutti}}]{guillochon17}
{Guillochon}, J., {Parrent}, J., {Kelley}, L.~Z., \& {Margutti}, R. 2017, \apj,
  835, 64

\bibitem[{{Heger} \& {Langer}(1998)}]{heger98}
{Heger}, A. \& {Langer}, N. 1998, \aap, 334, 210

\bibitem[{{Huang} {et~al.}(2015){Huang}, {Wang}, {Zhang}, {Brown}, {Zampieri},
  {Pumo}, {Zhang}, {Chen}, {Mo}, \& {Zhao}}]{huang15}
{Huang}, F., {Wang}, X., {Zhang}, J., {et~al.} 2015, \apj, 807, 59

\bibitem[{{Hubov{\'a}} \& {Pejcha}(2019)}]{hubova19}
{Hubov{\'a}}, D. \& {Pejcha}, O. 2019, \mnras, 489, 891

\bibitem[{{Inserra} {et~al.}(2011){Inserra}, {Turatto}, {Pastorello},
  {Benetti}, {Cappellaro}, {Pumo}, {Zampieri}, {Agnoletto}, {Bufano},
  {Botticella}, {Della Valle}, {Elias Rosa}, {Iijima}, {Spiro}, \&
  {Valenti}}]{inserra11}
{Inserra}, C., {Turatto}, M., {Pastorello}, A., {et~al.} 2011, \mnras, 417, 261

\bibitem[{{Jerkstrand}(2017)}]{jerkstrand17}
{Jerkstrand}, A. 2017, {Spectra of Supernovae in the Nebular Phase}, ed. A.~W.
  {Alsabti} \& P.~{Murdin}, 795

\bibitem[{{Kasen}(2010)}]{2010ApJ...708.1025K}
{Kasen}, D. 2010, \apj, 708, 1025

\bibitem[{{Kasen} {et~al.}(2017){Kasen}, {Metzger}, {Barnes}, {Quataert}, \&
  {Ramirez-Ruiz}}]{kasen17}
{Kasen}, D., {Metzger}, B., {Barnes}, J., {Quataert}, E., \& {Ramirez-Ruiz}, E.
  2017, \nat, 551, 80

\bibitem[{{Kashi} \& {Soker}(2010)}]{kashi10}
{Kashi}, A. \& {Soker}, N. 2010, \apj, 723, 602

\bibitem[{{Kee} {et~al.}(2014){Kee}, {Owocki}, \& {ud-Doula}}]{kee14}
{Kee}, N.~D., {Owocki}, S., \& {ud-Doula}, A. 2014, \mnras, 438, 3557

\bibitem[{{Kochanek}(2019)}]{kochanek19}
{Kochanek}, C.~S. 2019, \mnras, 483, 3762

\bibitem[{{Kokubo} {et~al.}(2019){Kokubo}, {Mitsuda}, {Morokuma}, {Tominaga},
  {Tanaka}, {Moriya}, {Yoachim}, {Ivezi{\'c}}, {Sako}, \& {Doi}}]{kokubo19}
{Kokubo}, M., {Mitsuda}, K., {Morokuma}, T., {et~al.} 2019, \apj, 872, 135

\bibitem[{{Kraus} {et~al.}(2007){Kraus}, {Borges Fernandes}, \& {de
  Ara{\'u}jo}}]{2007A&A...463..627K}
{Kraus}, M., {Borges Fernandes}, M., \& {de Ara{\'u}jo}, F.~X. 2007, \aap, 463,
  627

\bibitem[{{Krti{\v c}ka} {et~al.}(2011){Krti{\v c}ka}, {Owocki}, \&
  {Meynet}}]{2011A&A...527A..84K}
{Krti{\v c}ka}, J., {Owocki}, S.~P., \& {Meynet}, G. 2011, \aap, 527, A84

\bibitem[{{Kurf{\"u}rst} {et~al.}(2014){Kurf{\"u}rst}, {Feldmeier}, \& {Krti{\v
  c}ka}}]{2014A&A...569A..23K}
{Kurf{\"u}rst}, P., {Feldmeier}, A., \& {Krti{\v c}ka}, J. 2014, \aap, 569, A23

\bibitem[{{Kurf{\"u}rst} {et~al.}(2017){Kurf{\"u}rst}, {Feldmeier}, \& {Krti{\v
  c}ka}}]{2017ASPC..508...17K}
{Kurf{\"u}rst}, P., {Feldmeier}, A., \& {Krti{\v c}ka}, J. 2017, in
  Astronomical Society of the Pacific Conference Series, Vol. 508, The B[e]
  Phenomenon: Forty Years of Studies, ed. A.~{Miroshnichenko}, S.~{Zharikov},
  D.~{Kor{\v c}{\'a}kov{\'a}}, \& M.~{Wolf}, 17

\bibitem[{{Kurf{\"u}rst} {et~al.}(2018){Kurf{\"u}rst}, {Feldmeier}, \& {Krti{\v
  c}ka}}]{2018A&A...613A..75K}
{Kurf{\"u}rst}, P., {Feldmeier}, A., \& {Krti{\v c}ka}, J. 2018, \aap, 613, A75

\bibitem[{{Kurf{\"u}rst} \& {Krti{\v c}ka}(2017)}]{KKapplmath}
{Kurf{\"u}rst}, P. \& {Krti{\v c}ka}, J. 2017, Applications of Mathematics, 62,
  633

\bibitem[{{Kurf{\"u}rst} \& {Krti{\v c}ka}(2019)}]{kurfurst19}
{Kurf{\"u}rst}, P. \& {Krti{\v c}ka}, J. 2019, \aap, 625, A24

\bibitem[{{Leonard} {et~al.}(2000){Leonard}, {Filippenko}, {Barth}, \&
  {Matheson}}]{leonard00}
{Leonard}, D.~C., {Filippenko}, A.~V., {Barth}, A.~J., \& {Matheson}, T. 2000,
  \apj, 536, 239

\bibitem[{{Li} {et~al.}(2017){Li}, {Metzger}, {Chomiuk}, {Vurm}, {Strader},
  {Finzell}, {Beloborodov}, {Nelson}, {Shappee}, {Kochanek}, {Prieto}, {Kafka},
  {Holoien}, {Thompson}, {Luckas}, \& {Itoh}}]{li17}
{Li}, K.-L., {Metzger}, B.~D., {Chomiuk}, L., {et~al.} 2017, Nature Astronomy,
  1, 697

\bibitem[{{Lucy} {et~al.}(1989){Lucy}, {Danziger}, {Gouiffes}, \&
  {Bouchet}}]{lucy89}
{Lucy}, L.~B., {Danziger}, I.~J., {Gouiffes}, C., \& {Bouchet}, P. 1989, {Dust
  Condensation in the Ejecta of SN 1987 A}, ed. G.~{Tenorio-Tagle}, M.~{Moles},
  \& J.~{Melnick}, Vol. 350, 164

\bibitem[{{Mackey} {et~al.}(2014){Mackey}, {Mohamed}, {Gvaramadze}, {Kotak},
  {Langer}, {Meyer}, {Moriya}, \& {Neilson}}]{mackey14}
{Mackey}, J., {Mohamed}, S., {Gvaramadze}, V.~V., {et~al.} 2014, \nat, 512, 282

\bibitem[{{MacLeod} {et~al.}(2018){MacLeod}, {Ostriker}, \&
  {Stone}}]{macleod18}
{MacLeod}, M., {Ostriker}, E.~C., \& {Stone}, J.~M. 2018, \apj, 868, 136

\bibitem[{{Margutti} {et~al.}(2019){Margutti}, {Metzger}, {Chornock}, {Vurm},
  {Roth}, {Grefenstette}, {Savchenko}, {Cartier}, {Steiner}, {Terreran},
  {Margalit}, {Migliori}, {Milisavljevic}, {Alexand er}, {Bietenholz},
  {Blanchard}, {Bozzo}, {Brethauer}, {Chilingarian}, {Coppejans}, {Ducci},
  {Ferrigno}, {Fong}, {G{\"o}tz}, {Guidorzi}, {Hajela}, {Hurley}, {Kuulkers},
  {Laurent}, {Mereghetti}, {Nicholl}, {Patnaude}, {Ubertini}, {Banovetz},
  {Bartel}, {Berger}, {Coughlin}, {Eftekhari}, {Frederiks}, {Kozlova},
  {Laskar}, {Svinkin}, {Drout}, {MacFadyen}, \& {Paterson}}]{margutti19}
{Margutti}, R., {Metzger}, B.~D., {Chornock}, R., {et~al.} 2019, \apj, 872, 18

\bibitem[{{Matzner} \& {McKee}(1999)}]{1999ApJ...510..379M}
{Matzner}, C.~D. \& {McKee}, C.~F. 1999, \apj, 510, 379

\bibitem[{{Mauerhan} {et~al.}(2013){Mauerhan}, {Smith}, {Silverman},
  {Filippenko}, {Morgan}, {Cenko}, {Ganeshalingam}, {Clubb}, {Bloom},
  {Matheson}, \& {Milne}}]{mauerhan13}
{Mauerhan}, J.~C., {Smith}, N., {Silverman}, J.~M., {et~al.} 2013, \mnras, 431,
  2599

\bibitem[{{Mauerhan} {et~al.}(2017){Mauerhan}, {Van Dyk}, {Johansson}, {Hu},
  {Fox}, {Wang}, {Graham}, {Filippenko}, \& {Shivvers}}]{mauerhan17}
{Mauerhan}, J.~C., {Van Dyk}, S.~D., {Johansson}, J., {et~al.} 2017, \apj, 834,
  118

\bibitem[{{McDowell} {et~al.}(2018){McDowell}, {Duffell}, \&
  {Kasen}}]{mcdowell18}
{McDowell}, A.~T., {Duffell}, P.~C., \& {Kasen}, D. 2018, \apj, 856, 29

\bibitem[{{Metzger} \& {Pejcha}(2017)}]{metzger17}
{Metzger}, B.~D. \& {Pejcha}, O. 2017, \mnras, 471, 3200

\bibitem[{{Moe} \& {Di Stefano}(2017)}]{moe17}
{Moe}, M. \& {Di Stefano}, R. 2017, \apjs, 230, 15

\bibitem[{{Moriya} {et~al.}(2013){Moriya}, {Blinnikov}, {Baklanov}, {Sorokina},
  \& {Dolgov}}]{moriya13}
{Moriya}, T.~J., {Blinnikov}, S.~I., {Baklanov}, P.~V., {Sorokina}, E.~I., \&
  {Dolgov}, A.~D. 2013, \mnras, 430, 1402

\bibitem[{{Moriya} {et~al.}(2018){Moriya}, {F{\"o}rster}, {Yoon},
  {Gr{\"a}fener}, \& {Blinnikov}}]{moriya18}
{Moriya}, T.~J., {F{\"o}rster}, F., {Yoon}, S.-C., {Gr{\"a}fener}, G., \&
  {Blinnikov}, S.~I. 2018, \mnras, 476, 2840

\bibitem[{{Morozova} {et~al.}(2015){Morozova}, {Piro}, {Renzo}, {Ott},
  {Clausen}, {Couch}, {Ellis}, \& {Roberts}}]{2015ApJ...814...63M}
{Morozova}, V., {Piro}, A.~L., {Renzo}, M., {et~al.} 2015, \apj, 814, 63

\bibitem[{{Morris} \& {Podsiadlowski}(2007)}]{morris07}
{Morris}, T. \& {Podsiadlowski}, P. 2007, Science, 315, 1103

\bibitem[{{M{\"u}ller} {et~al.}(2017){M{\"u}ller}, {Prieto}, {Pejcha}, \&
  {Clocchiatti}}]{2017ApJ...841..127M}
{M{\"u}ller}, T., {Prieto}, J.~L., {Pejcha}, O., \& {Clocchiatti}, A. 2017,
  \apj, 841, 127

\bibitem[{{Nyholm} {et~al.}(2017){Nyholm}, {Sollerman}, {Taddia}, {Fremling},
  {Moriya}, {Ofek}, {Gal-Yam}, {De Cia}, {Roy}, {Kasliwal}, {Cao}, {Nugent}, \&
  {Masci}}]{2017A&A...605A...6N}
{Nyholm}, A., {Sollerman}, J., {Taddia}, F., {et~al.} 2017, \aap, 605, A6

\bibitem[{{Okazaki}(2001)}]{2001PASJ...53..119O}
{Okazaki}, A.~T. 2001, \pasj, 53, 119

\bibitem[{{Pandolfi} \& {D'Ambrosio}(2001)}]{2001JCoPh.166..271P}
{Pandolfi}, M. \& {D'Ambrosio}, D. 2001, Journal of Computational Physics, 166,
  271

\bibitem[{{Patat} {et~al.}(2011){Patat}, {Taubenberger}, {Benetti},
  {Pastorello}, \& {Harutyunyan}}]{patat11}
{Patat}, F., {Taubenberger}, S., {Benetti}, S., {Pastorello}, A., \&
  {Harutyunyan}, A. 2011, \aap, 527, L6

\bibitem[{{Paxton} {et~al.}(2011){Paxton}, {Bildsten}, {Dotter}, {Herwig},
  {Lesaffre}, \& {Timmes}}]{paxton11}
{Paxton}, B., {Bildsten}, L., {Dotter}, A., {et~al.} 2011, \apjs, 192, 3

\bibitem[{{Paxton} {et~al.}(2013){Paxton}, {Cantiello}, {Arras}, {Bildsten},
  {Brown}, {Dotter}, {Mankovich}, {Montgomery}, {Stello}, {Timmes}, \&
  {Townsend}}]{paxton13}
{Paxton}, B., {Cantiello}, M., {Arras}, P., {et~al.} 2013, \apjs, 208, 4

\bibitem[{{Pejcha} {et~al.}(2016{\natexlab{a}}){Pejcha}, {Metzger}, \&
  {Tomida}}]{pejcha16b}
{Pejcha}, O., {Metzger}, B.~D., \& {Tomida}, K. 2016{\natexlab{a}}, \mnras,
  461, 2527

\bibitem[{{Pejcha} {et~al.}(2016{\natexlab{b}}){Pejcha}, {Metzger}, \&
  {Tomida}}]{pejcha16a}
{Pejcha}, O., {Metzger}, B.~D., \& {Tomida}, K. 2016{\natexlab{b}}, \mnras,
  455, 4351

\bibitem[{{Pejcha} {et~al.}(2017){Pejcha}, {Metzger}, {Tyles}, \&
  {Tomida}}]{pejcha17}
{Pejcha}, O., {Metzger}, B.~D., {Tyles}, J.~G., \& {Tomida}, K. 2017, \apj,
  850, 59

\bibitem[{{Pejcha} \& {Prieto}(2015)}]{pejcha15}
{Pejcha}, O. \& {Prieto}, J.~L. 2015, \apj, 806, 225

\bibitem[{{Podsiadlowski} {et~al.}(1992){Podsiadlowski}, {Joss}, \&
  {Hsu}}]{podsiadlowski92}
{Podsiadlowski}, P., {Joss}, P.~C., \& {Hsu}, J.~J.~L. 1992, \apj, 391, 246

\bibitem[{{Sana} {et~al.}(2012){Sana}, {de Mink}, {de Koter}, {Langer},
  {Evans}, {Gieles}, {Gosset}, {Izzard}, {Le Bouquin}, \& {Schneider}}]{sana12}
{Sana}, H., {de Mink}, S.~E., {de Koter}, A., {et~al.} 2012, Science, 337, 444

\bibitem[{{Smith}(2011)}]{smith11}
{Smith}, N. 2011, \mnras, 415, 2020

\bibitem[{{Smith}(2013{\natexlab{a}})}]{smith13_etacar}
{Smith}, N. 2013{\natexlab{a}}, \mnras, 429, 2366

\bibitem[{{Smith}(2013{\natexlab{b}})}]{smith13}
{Smith}, N. 2013{\natexlab{b}}, \mnras, 434, 102

\bibitem[{{Smith}(2014)}]{smith14}
{Smith}, N. 2014, \araa, 52, 487

\bibitem[{{Smith}(2017)}]{smith17_handbook}
{Smith}, N. 2017, {Interacting Supernovae: Types IIn and Ibn}, ed. A.~W.
  {Alsabti} \& P.~{Murdin}, 403

\bibitem[{{Smith} {et~al.}(2016){Smith}, {Andrews}, {Mauerhan}, {Zheng},
  {Filippenko}, {Graham}, \& {Milne}}]{smith16}
{Smith}, N., {Andrews}, J.~E., {Mauerhan}, J.~C., {et~al.} 2016, \mnras, 455,
  3546

\bibitem[{{Smith} {et~al.}(2018){Smith}, {Andrews}, {Rest}, {Bianco}, {Prieto},
  {Matheson}, {James}, {Smith}, {Strampelli}, \& {Zenteno}}]{smith18}
{Smith}, N., {Andrews}, J.~E., {Rest}, A., {et~al.} 2018, \mnras, 480, 1466

\bibitem[{{Smith} {et~al.}(2008){Smith}, {Foley}, \& {Filippenko}}]{smith08}
{Smith}, N., {Foley}, R.~J., \& {Filippenko}, A.~V. 2008, \apj, 680, 568

\bibitem[{{Smith} {et~al.}(2015){Smith}, {Mauerhan}, {Cenko}, {Kasliwal},
  {Silverman}, {Filippenko}, {Gal-Yam}, {Clubb}, {Graham}, {Leonard}, {Horst},
  {Williams}, {Andrews}, {Kulkarni}, {Nugent}, {Sullivan}, {Maguire}, {Xu}, \&
  {Ben-Ami}}]{smith15}
{Smith}, N., {Mauerhan}, J.~C., {Cenko}, S.~B., {et~al.} 2015, \mnras, 449,
  1876

\bibitem[{{Smith} \& {McCray}(2007)}]{smith07}
{Smith}, N. \& {McCray}, R. 2007, \apjl, 671, L17

\bibitem[{{Smith} {et~al.}(2009){Smith}, {Silverman}, {Chornock}, {Filippenko},
  {Wang}, {Li}, {Ganeshalingam}, {Foley}, {Rex}, \& {Steele}}]{smith09}
{Smith}, N., {Silverman}, J.~M., {Chornock}, R., {et~al.} 2009, \apj, 695, 1334

\bibitem[{{Steinberg} \& {Metzger}(2018)}]{steinberg18}
{Steinberg}, E. \& {Metzger}, B.~D. 2018, \mnras, 479, 687

\bibitem[{{Stevens} {et~al.}(1992){Stevens}, {Blondin}, \&
  {Pollock}}]{stevens92}
{Stevens}, I.~R., {Blondin}, J.~M., \& {Pollock}, A.~M.~T. 1992, \apj, 386, 265

\bibitem[{{Stritzinger} {et~al.}(2012){Stritzinger}, {Taddia}, {Fransson},
  {Fox}, {Morrell}, {Phillips}, {Sollerman}, {Anderson}, {Boldt}, {Brown},
  {Campillay}, {Castellon}, {Contreras}, {Folatelli}, {Habergham}, {Hamuy},
  {Hjorth}, {James}, {Krzeminski}, {Mattila}, {Persson}, \&
  {Roth}}]{stritzinger12}
{Stritzinger}, M., {Taddia}, F., {Fransson}, C., {et~al.} 2012, \apj, 756, 173

\bibitem[{{Sugerman} {et~al.}(2006){Sugerman}, {Ercolano}, {Barlow}, {Tielens},
  {Clayton}, {Zijlstra}, {Meixner}, {Speck}, {Gledhill}, {Panagia}, {Cohen},
  {Gordon}, {Meyer}, {Fabbri}, {Bowey}, {Welch}, {Regan}, \&
  {Kennicutt}}]{sugerman06}
{Sugerman}, B. E.~K., {Ercolano}, B., {Barlow}, M.~J., {et~al.} 2006, Science,
  313, 196

\bibitem[{{Suzuki} {et~al.}(2019){Suzuki}, {Moriya}, \& {Takiwaki}}]{suzuki19}
{Suzuki}, A., {Moriya}, T.~J., \& {Takiwaki}, T. 2019, \apj, 887, 249

\bibitem[{{Szalai} {et~al.}(2019){Szalai}, {Vink{\'o}}, {K{\"o}nyves-T{\'o}th},
  {Nagy}, {Bostroem}, {S{\'a}rneczky}, {Brown}, {Pejcha}, {B{\'o}di}, {Cseh},
  {Cs{\"o}rnyei}, {Dencs}, {Hanyecz}, {Ign{\'a}cz}, {Kalup}, {Kriskovics},
  {Ordasi}, {P{\'a}l}, {Seli}, {S{\'o}dor}, {Szak{\'a}ts}, {Vida}, {Zsidi},
  {Konkoly Team}, {Arcavi}, {Ashall}, {Burke}, {Galbany}, {Hiramatsu},
  {Hosseinzadeh}, {Hsiao}, {Howell}, {McCully}, {Moran}, {Rho}, {Sand },
  {Shahbandeh}, {Valenti}, {Wang}, {Wheeler}, \& {Supernova
  Project}}]{szalai19}
{Szalai}, T., {Vink{\'o}}, J., {K{\"o}nyves-T{\'o}th}, R., {et~al.} 2019, \apj,
  876, 19

\bibitem[{{Taddia} {et~al.}(2020){Taddia}, {Stritzinger}, {Fransson}, {Brown},
  {Contreras}, {Holmbo}, {Moriya}, {Phillips}, {Sollerman}, {Suntzeff},
  {Ashall}, {Burns}, {Busta}, {Campillay}, {Castell{\'o}n}, {Corco}, {Di
  Mille}, {Gall}, {Gonz{\'a}lez}, {Hsiao}, {Morrell}, {Nyholm}, {Simon}, \&
  {Ser{\'o}n}}]{taddia20}
{Taddia}, F., {Stritzinger}, M.~D., {Fransson}, C., {et~al.} 2020, \aap, 638,
  A92

\bibitem[{{Terreran} {et~al.}(2016){Terreran}, {Jerkstrand}, {Benetti},
  {Smartt}, {Ochner}, {Tomasella}, {Howell}, {Morales-Garoffolo},
  {Harutyunyan}, {Kankare}, {Arcavi}, {Cappellaro}, {Elias-Rosa},
  {Hosseinzadeh}, {Kangas}, {Pastorello}, {Tartaglia}, {Turatto}, {Valenti},
  {Wiggins}, \& {Yuan}}]{terreran16}
{Terreran}, G., {Jerkstrand}, A., {Benetti}, S., {et~al.} 2016, \mnras, 462,
  137

\bibitem[{{van Marle} {et~al.}(2010){van Marle}, {Smith}, {Owocki}, \& {van
  Veelen}}]{vanmarle10}
{van Marle}, A.~J., {Smith}, N., {Owocki}, S.~P., \& {van Veelen}, B. 2010,
  \mnras, 407, 2305

\bibitem[{{Vishniac}(1994)}]{vishniac94}
{Vishniac}, E.~T. 1994, \apj, 428, 186

\bibitem[{{Vlasis} {et~al.}(2016){Vlasis}, {Dessart}, \& {Audit}}]{vlasis16}
{Vlasis}, A., {Dessart}, L., \& {Audit}, E. 2016, \mnras, 458, 1253

\bibitem[{{Wang} \& {Wheeler}(2008)}]{wang08}
{Wang}, L. \& {Wheeler}, J.~C. 2008, \araa, 46, 433

\bibitem[{{Weil} {et~al.}(2020){Weil}, {Fesen}, {Patnaude}, \&
  {Milisavljevic}}]{weil20}
{Weil}, K.~E., {Fesen}, R.~A., {Patnaude}, D.~J., \& {Milisavljevic}, D. 2020,
  arXiv e-prints, arXiv:2006.02496

\bibitem[{{Wilkin}(1996)}]{1996ApJ...459L..31W}
{Wilkin}, F.~P. 1996, \apjl, 459, L31

\bibitem[{{Yuan} {et~al.}(2016){Yuan}, {Jerkstrand}, {Valenti}, {Sollerman},
  {Seitenzahl}, {Pastorello}, {Schulze}, {Chen}, {Childress}, {Fraser},
  {Fremling}, {Kotak}, {Ruiter}, {Schmidt}, {Smartt}, {Taddia}, {Terreran},
  {Tucker}, {Barbarino}, {Benetti}, {Elias-Rosa}, {Gal-Yam}, {Howell},
  {Inserra}, {Kankare}, {Lee}, {Li}, {Maguire}, {Margheim}, {Mehner}, {Ochner},
  {Sullivan}, {Tomasella}, \& {Young}}]{yuan16}
{Yuan}, F., {Jerkstrand}, A., {Valenti}, S., {et~al.} 2016, \mnras, 461, 2003

\bibitem[{{Zel'dovich} \& {Raizer}(1967)}]{1967pswh.book.....Z}
{Zel'dovich}, {\relax Ya}.~B. \& {Raizer}, {\relax Yu}.~P. 1967, {Physics of
  shock waves and high-temperature hydrodynamic phenomena}

\end{thebibliography}

\begin{appendix}
\onecolumn
\section{Analytical scaling of SN-wind and SN-disk interaction}
\label{analsol}
Here we summarize the equations that are used to analytically estimate the power produced 
by a strong shock that propagates through the stellar wind or through analytically scaled 
CSM configurations like a circumstellar disk. Results of these estimates were used in 
Sect.~\ref{lightcurves} and Fig.~\ref{fig:lc}. 
We start by constructing analytic description of the density profiles of the SN ejecta 
and the CSM. For SN ejecta, we follow the broken power law defined by \citet{chevalier89} 
with an inner flatter region and an outer steeper region. The distance of the transition 
point between the two regions and the velocity at this point are related as 
$R_\text{tr}=\varv_\text{tr} t$. The density of a spherically symmetric ejecta 
is $\rho_\text{ej,in}(r,t)=\rho_\text{tr}\left(R_\text{tr}/r\right)^{\delta}$ 
for $r<R_\text{tr}$, $\rho_\text{ej,out}(r,t)=\rho_\text{tr}\left(R_\text{tr}/r\right)^{n}$ 
for $R_\text{tr}<r<R_\text{SN}$, where $R_\text{SN}=\varv_\text{max}\,t$ is the velocity 
of the outermost layer of the expanding SN ejecta. The density and velocity of the transition 
point are 
\begin{align}
\label{Mori02}
\rho_\text{tr}=\frac{(3-\delta)(n-3)}{4\pi\,(n-\delta)} 
\frac{M_\text{ej}}{R_\text{tr}^{3}},\quad \varv_\text{tr}=
\left[\frac{2(5-\delta)(n-5)}{(3-\delta)(n-3)} \frac{E_\text{ej}}{M_\text{ej}}\right]^{1/2},
\end{align}
where $M_\text{ej}$ and $E_\text{ej}$ are mass and energy of SN ejecta, respectively, 
while $\rho_\text{ej}=0$ for $r>R_\text{SN}$. The inner density slope 
$\delta\simeq 0\text{ - }1$ and the outer density slope $n\simeq 10$ is expected 
for SNe Ib/Ic and SNe Ia
\citep[][]{1999ApJ...510..379M,2010ApJ...708.1025K} while $n\simeq 12$ is commonly 
accepted for RSGs \citep[][]{1999ApJ...510..379M}.

We describe the density profile $\rho_\text{sw}(r)$ of spherically symmetric  
circumstellar environment (wind) in Eq.~\eqref{windpower0} and the density profile 
$\rho_\text{disk}(r,\theta)$ of the thin circumstellar disk in Eq.~\eqref{rhocsm} 
\citep[cf.][cf. also Eq.~(37) in \citealt{2018A&A...613A..75K}]{kurfurst19}. 
The total mass of the shocked SN ejecta and CSM, $M_\text{sh}$, becomes
\begin{align}
\label{mass1}
M_\text{sh}= \int_{R_\text{sh}}^{R_\text{SN}}4\pi r^2\rho_\text{ej}\,\text{d}r +
\int_{R_\star}^{R_\text{sh}}4\pi r^2\left(\rho_\text{sw}+\rho_\text{disk}\right)\text{d}r,
\end{align}
where $R_\text{sh}=\varv_\text{sh}t$ is the radius of the shock front. We integrate the 
second term in the right-hand side of Eq.~\eqref{mass1} from the stellar radius 
$R_\star\ll R_\text{sh}$, because we assume the CSM is sweeped up to the radius $R_\text{sh}$,
forming there a thin shell. We obtain
\begin{align}\label{mass2}
\frac{M_\text{sh}}{4\pi}= \frac{B_\text{sn}}{n-3}\left(\frac{t}{R_\text{sh}}\right)^{n-3}+
\frac{D_\text{sw}}{3-s}R_\text{sh}^{3-s}+ \frac{D_\text{disk}}{3-w}R_\text{sh}^{3-w},
\end{align}
where the constants 
$B_\text{sn}=\rho_\text{tr}\,\varv_\text{tr}^n\,t^3$, $D_\text{sw}=\rho_{0,\text{wind}} 
R_\star^s$ and $D_\text{disk}=\rho_{0,\text{disk}} R_\star^w$, $s$ and $w$ are pre-explosion 
density slopes of wind and disk, respectively; for the meaning of base densities 
$\rho_{0,\text{wind}}$ and $\rho_{0,\text{disk}}$, see Sect.~\ref{intro} and 
Eq.~\eqref{rhocsm} in Sect.~\ref{numset}.

Employing the shocked shell pressure force (scaled from Rankine-Hugoniot relations), 
assuming a strong shock and neglecting pre-explosion velocities of CSM,
\begin{align}\label{mass3}
 \left[\rho_\text{ej}\left(\varv_\text{SN}-\varv_\text{sh}\right)^2- 
 \left(\rho_\text{sw}+\rho_\text{disk}\right)\varv_\text{sh}^2\right]4\pi R_\text{sh}^2 =
 M_\text{sh}\frac{\text{d}\varv_\text{sh}}{\text{d}t},
\end{align}
and by substituting Eq.~\eqref{mass2} and the shock velocity into Eq.~\eqref{mass3}, 
we obtain
\begin{align}\label{mass4}
 \frac{B_\text{sn}t^{n-3}}{R_\text{sh}^{n-2}}\left(\frac{R_\text{sh}}{t}- 
 \frac{\text{d}R_\text{sh}}{\text{d}t}\right)^2 - \left(D_\text{sw} R_\text{sh}^{2-s}+ 
 D_\text{disk}R_\text{sh}^{2-w} \right)\left(\frac{\text{d}R_\text{sh}}{\text{d}t}\right)^2=
 \left[ \frac{B_\text{sn}}{n-3}\left(\frac{t}{R_\text{sh}}\right)^{n-3}+ 
 \frac{D_\text{sw}R_\text{sh}^{3-s}}{3-s}+ \frac{D_\text{disk}R_\text{sh}^{3-w}}{3-w}\right] 
 \frac{\text{d}^2R_\text{sh}}{\text{d}t^2}.
\end{align}
Assuming the solution of Eq.~\eqref{mass4} as a power law in the form 
$R_\text{sh}(t)=At^\alpha$, we analytically evaluate the factor $A$ for a shock wave 
that propagates through CSM. We write the explicit form of the power law solution 
for $R_\text{sh,\,sw}$ and $R_\text{sh,\,disk}$, respectively, as
\begin{align}\label{mass6}
 R_\text{sh,\,sw}(t)=\left[\frac{(3-s)(4-s)}{(n-3)(n-4)}  \frac{B_\text{sn}}
 {D_\text{sw}}\right]^{1/(n-s)}t^{(n-3)/(n-s)}, \qquad
 R_\text{sh,\,disk}(t)=\left[\frac{\left(\widetilde{H}+3-w\right)(3-w)}{(n-3)(n-4)} 
 \frac{B_\text{sn}}{D_\text{disk}}\right]^{1/(n-w)}t^{(n-3)/(n-w)},
\end{align}
where $\widetilde{H} = H/\mathcal{R}$ (Eq.~[\ref{rhocsm}]). If we substitute 
$n=12,\,s=2,\,w=2$, and $\widetilde{H}=0.2$, we get 
$R_\text{sh,\,sw}(t)\approx 0.7\left(B_\text{sn}/D_\text{sw}\right)^{1/10}\,t^{9/10}$ and 
$R_\text{sh,\,disk}(t)\approx 0.66\left(B_\text{sn}/D_\text{disk}\right)^{1/10}\,t^{9/10}$. 
We use here the approximation $\widetilde{H}=\text{const.}$ for analytical feasibility, 
even if $\widetilde{H}$ is not a constant in general unless the disk temperature radially 
decreases as $T\sim r^{-1}$ \citep[cf.][]{2014A&A...569A..23K,2018A&A...613A..75K}.

We estimate the shock power as the product of pressure force exerted by the 
shock on the unshocked material and the shock velocity \citep[cf.][]{mcdowell18},
\begin{equation}
\label{power1}
\dot{Q}_\text{sw}=P_\text{sh}\,\varv_\text{sh}\,S_\text{sh,sw},\quad 
\dot{Q}_\text{disk}=P_\text{sh}\,\varv_\text{sh}\,S_\text{sh,disk}
\end{equation}
where $P_\text{sh}$ is the pressure behind the shock front that propagates through 
the stellar wind or the disk region, $\varv_\text{sh}$ is the corresponding shock 
velocity (that can be derived, e.g., from second Equation~\eqref{mass6}), 
and $S_\text{sh}$ is the surface area of the shock in the wind or disk region. 
For a strong shock is the pressure $P_\text{sh}$ given by \citep{1967pswh.book.....Z} 
\begin{equation}
\label{power2}
P_\text{sh}=\frac{2}{\gamma-1}\,\rho\,\varv_\text{sh}^2,
\end{equation}
where $\rho$ is either $\rho_\text{wind}$ (Eq.~[\ref{windpower0}]) 
or $\rho_\text{disk}$ (Eq.~[\ref{rhocsm}]), simplified for the case of thin equatorial 
disk, $\theta\approx\pi/2$. We can express the corresponding shock surface area simply as 
\begin{equation}\label{power3}
S_\text{sh,sw}=4\pi R_\text{sh}^2,\quad S_\text{sh,disk}=4\pi \widetilde{H}R_\text{sh}^2,
\end{equation}
respectively. 

Following these assumptions, we obtain the analytically estimated shock heating rate 
(shock power) for the disk angular region as
\begin{equation}\label{power5}
\dot{Q}_\text{disk}=\left(\frac{n-3}{n-w}\right)^3
\frac{8\pi \widetilde{H}D_\text{disk}}{\gamma-1}\,
\mathcal{A}_\text{disk}^{(5-w)/(n-w)}\,t^{(5-w)(n-3)/(n-w)-3},
\end{equation}
where $\mathcal{A_\text{disk}}$ is the factor in square brackets in second 
Equation~\eqref{mass6}. The shock heating rate $\dot{Q}_\text{sw}$ for the spherically 
symmetric wind region is quite analogous; we may set a simple  
approximation $1-\widetilde{H}\approx 1$, or we can calculate $\dot{Q}_\text{sw}$ 
using Eq.~\eqref{windpower0}. We also estimate the ratio 
$\dot{Q}_\text{disk}/\dot{Q}_\text{sw}$, that is, what will be the contribution of the 
disk to the shock powered luminosity to that of the whole spherically symmetric CSM. 
If, for example, the parameters 
$s= 2$, $w= 2$, $n= 7\text{ - }12$, $\widetilde{H}= 0.05\text{ - }0.2$ 
(see Eq.~[\ref{mass6}] for the meaning of the parameters), we obtain the 
ratio $\dot{Q}_\text{disk}/\dot{Q}_\text{sw}\sim 10^3\text{ - }10^4$.

\end{appendix}

\end{document}